\documentclass{emulateapj}

\usepackage{epstopdf}
\usepackage{amsmath}

\newcommand{\tableskip}{\\[-6pt]}

\shorttitle{Detecting Baryon Acoustic Oscillations}
\shortauthors{Labatie et al.}

\begin{document}


\title{Detecting Baryon Acoustic Oscillations}


\author{A. Labatie \altaffilmark{1} and J.L. Starck}
\affil{Laboratoire AIM (UMR 7158), CEA/DSM-CNRS-Universit\'e Paris Diderot, IRFU, SEDI-SAP, Service d'Astrophysique, 
   	Centre de Saclay, F-91191 Gif-Sur-Yvette cedex, France}
\email{antoine.labatie@cea.fr}
\author{M. Lachi\`eze-Rey}
\affil{Astroparticule et Cosmologie (APC), CNRS-UMR 7164, Universit\'e Paris 7 Denis Diderot, 
            10, rue Alice Domon et L\'eonie Duquet F-75205 Paris Cedex 13, France}



\begin{abstract}
Baryon Acoustic Oscillations (BAOs) are a feature imprinted in the galaxy distribution by acoustic waves traveling in the plasma of the early universe. Their detection at the expected scale in large-scale structures strongly supports current cosmological models with a nearly linear evolution from redshift $z \approx 1000$ and the existence of dark energy. Besides, BAOs provide a standard ruler for studying cosmic expansion. In this paper we focus on methods for BAO detection using the correlation function measurement $\hat{\xi}$. For each method, we want to understand the tested hypothesis (the hypothesis $\mathcal{H}_0$ to be rejected) and the underlying assumptions. We first present wavelet methods which are mildly model-dependent and mostly sensitive to the BAO feature. Then we turn to fully model-dependent methods. We present the most often used method based on the $\chi^2$ statistic, but we find it has limitations. In general the assumptions of the $\chi^2$ method are not verified, and it only gives a rough estimate of the significance. The estimate can become very wrong when considering more realistic hypotheses, where the covariance matrix of $\hat{\xi}$ depends on cosmological parameters. Instead we propose to use the $\Delta l$ method based on two modifications: we modify the procedure for computing the significance and make it rigorous, and we modify the statistic to obtain better results in the case of varying covariance matrix. We verify with simulations that correct significances are different from the ones obtained using the classical $\chi^2$ procedure. We also test a simple example of varying covariance matrix. In this case we find that our modified statistic outperforms the classical $\chi^2$ statistic when both significances are correctly computed. Finally we find that taking into account variations of the covariance matrix can change both BAO detection levels and cosmological parameter constraints.
\end{abstract}





\section{Introduction} 

Large-scale structures in the Universe provide crucial information and can be used to test different cosmological models. This study is complementary to other observations such as the Cosmic Microwave Background (CMB) or Type Ia supernovae. Combining those different observations enables to cross-test models, to break degeneracies, and better constrain cosmological parameters. The good agreement found recently between them and the now-standard Lambda-Cold Dark Matter ($\Lambda\mbox{CDM}$) model gives hope for this model to be a lasting foundation. 

$\mbox{CDM}$ models with baryons predict the existence of acoustic waves traveling in the hot plasma before recombination, when baryons and photons were coupled together. Those spherical waves originate from the competition between gravitation making over-densities collapse and the photon pressure. About 380 000 years after the Big Bang, baryons and photons decoupled and those spherical waves became frozen at the sound horizon scale $r_s$. Because of their large size ($\approx 153$ Mpc) their imprints in the matter density field have mainly undertook linear evolution and they should be clearly seen in current large-scale structures. Those acoustic waves which translate into an excess of correlation at the sound horizon scale are known as Baryon Acoustic Oscillations (BAOs, \cite{Pee70,Bas10}).

There are two different uses of BAOs that should be distinguished. First they can be used as a very distinct feature to confirm cosmological models. Indeed their detection at the expected scale in large-scale structures gives a strong support for CDM models, with a linear gravitational evolution from $z \approx 1000$ and the existence of dark energy. Concretely, the detection is made through hypothesis testing, by finding that models with BAOs are strongly preferred to models without BAOs.

The first convincing detection of BAOs in large-scale structures was reported in the correlation function analysis  \citep{Eis05} of the Sloan Digital Sky Survey (SDSS, \cite{Yor00}) Luminous Red Galaxies (LRG, \cite{Eis01}) survey. It gave a $3.4\sigma$ significance in Data Release 3 (DR3). It was followed by other detections, as in the 2-degree Field Galaxy Redshift Survey (2dFGRS, \cite{Colless01}), using power spectrum analysis \citep{Cole05} with a $2.5\sigma$ significance. The power spectrum analysis is also applied in \cite{Hut06} with a $3.3\sigma$ detection in the SDSS-LRG DR4. In \cite{Per07} and \cite{Per10} the combined power spectrum of LRG and ÔMainÕ \citep{Str02} samples of SDSS (with respectively  DR5 and DR7), together with the 2dFGRS survey is used to obtain respective significances $3\sigma$ and $3.6\sigma$. Very recently BAOs were also detected in the power spectrum \citep{Bla11} of the WiggleZ Dark Energy Survey \citep{Dri10} at a higher redshift $z=0.6$ with a significance of $3.2 \sigma$.

We must keep in mind that the different detection levels cannot usually be compared. In \cite{Eis05} and \cite{Cole05} it is calculated with respect to zero-baryon models (pure CDM models). In \cite{Hut06} and \cite{Bla11}, the significance is calculated with respect to the "no-wiggles" fits of \cite{Eis98}, which remove the baryon oscillations signature but keep the intermediate suppression of power due to baryons. Finally in \cite{Per07} and \cite{Per10}, it is calculated with respect to power spectrum models where oscillations are smoothed out using splines. 

The second use of BAOs consists in constraining cosmological parameters. Again BAOs are very useful, because they provide a statistical standard ruler \citep{Seo03} with an absolute size given with small uncertainty by CMB measurements \citep{Kom09}. So they directly constrain the redshift-distance relation in redshift surveys. Besides, BAOs appear to have the lowest systematic uncertainties among current methods for studying cosmic expansion \citep{Alb06}. They have been used in combination with other cosmological probes to constrain more efficiently cosmological parameters \citep{Eis05,Teg06,Per07,San09,Per10,Rei10,Kaz10,Bla11}.

Note that in most studies, the aforementioned constraints not only come from BAOs, but from the whole information in the estimated correlation function or power spectrum. Note also that BAOs do not need to be detected before they can be used for parameter constraints \citep{Cab11}. One could think that the BAO peak must be proved to be significant and not a random fluctuation, before it is used as a standard ruler. However all sources of uncertainty are normally taken into account when obtaining the constraints (e.g. in the covariance matrix of $\hat{\xi}$ when using the correlation function), so this argument is not valid. The real question is whether the accepted cosmological models are correct, and BAO detection in a given sample is just a contribution to support these models. 

In this paper we focus on the first use of BAOs, i.e. on the BAO detection. We restrict the analysis to the correlation function, but most reasonings could also be applied to the power spectrum.

The plan of this paper is as follows: we start in section \ref{correlation_section} by discussing the correlation function estimation and modeling, as well as the general procedure for BAO detection. In section \ref{wavelet_section} we present wavelet methods for BAO detection, which are mildly model-dependent. In the rest of the paper we focus on fully model-dependent methods. In section \ref{chi2_section} we present the classical method used for BAO detection, based on the $\chi^2$ statistic. We find that this method does not provide the correct significance. So we propose in section \ref{modified_chi2_section} a new method that we call the $\Delta l$ method, based on two modifications to the $\chi^2$ method. In section \ref{constraints_section} we explain the other use of BAOs, i.e. how parameter constraints can be obtained. Finally we illustrate the different methods and results using simulations in section \ref{test_simu}.

\section{BAO detection in the correlation function}
\label{correlation_section}

\subsection{Correlation function}
The correlation function $\xi$ is a second order statistic that measures the clustering of a continuous field or point process. More precisely for the distribution of galaxies, it quantifies the excess of probability to find a pair of galaxies in volumes $dV_1$ and $dV_2$ separated by ${\bf r}$, compared to a random unclustered distribution

\begin{equation}
dP_{12}= \bar{n} \left[1+\xi({\bf r})\right] dV_1 dV_2
\end{equation}

with $\bar{n}$ the mean density of the point distribution. With the isotropy hypothesis, $\xi$ only depends on the norm $r=\| {\bf r} \|$ of the separation vector ${\bf r}$. However in redshift space the field is not rigorously isotropic. In this case, one is usually interested in the monopole $\xi(r)$ of $\xi({\bf r})$. In the following we will make the abuse of language of referring to the monopole when speaking about the correlation function.

Given a galaxy survey, the correlation function can be estimated by comparing the number of pairs at distance $r$ with a random catalogue in the same volume. Different estimators based on this method have been proposed and empirically compared \citep{Lab11,Ker00,Pon99}). While \cite{Pon99} did not recommend one estimator for all cases, \cite{Ker00} recommend to use the Landy-Szalay estimator. The recommendation is the same in the more recent study \cite{Lab11}, where Landy-Szalay is found to be nearly unbiased for current galaxy surveys. It is given by

\begin{equation}
\hat{\xi}(r) =  1 + {N_{RR} \over N_{DD}} {  DD(r) \over  RR(r)}  -  2 {N_{RR} \over N_{DR}}  { DR(r) \over RR(r)}
\end{equation}

with $DD(r)$, $RR(r)$, and $DR(r)$ the number of pairs at a distance in $[r \pm dr/2]$ of respectively data-data, random-random and data-random points, and $N_{DD}$, $N_{RR}$ and $N_{DR}$ the total number of corresponding pairs in the catalogues.

\subsection{Modeling the galaxy correlation function}
\label{correl_model}
In the early universe prior to recombination, baryons are tightly coupled with photons. This results in acoustic waves traveling in the plasma (BAOs), but also in the suppression of power on small and intermediate scales compared to a CDM model without baryons. After the time of decoupling (also called drag epoch),  the matter density field becomes pressureless, allowing the perturbations to grow by gravitational instability. This evolution can be analytically solved in the linear regime where fluctuations are small, and only the overall amplitude of the power spectrum is changing.

\cite{Eis98} provides fitting formulae for the linear power spectrum, with a dependence on cosmological parameters. From the linear power spectrum, the linear correlation function is simply obtained by Fourier transform. The effect of BAOs is clearly identified as series of wiggles in the power spectrum, and as a localized bump in the correlation function at the sound horizon scale $r_s$ (see figure \ref{cf_plot}).

To fully model the matter correlation function, one also has to take into account non-linear evolution. This can be done using $N$-body simulations, which empirically provide a correction from the linear to the non-linear correlation functions. \cite{Smi03} provide corrections for scale-free power spectrum. So one also has to correct for non-linear degradation of the acoustic peak. \cite{Eis07} found that this is well approximated by a Gaussian smoothing of the acoustic feature in real and redshift spaces.

A last step to model the galaxy correlation function is to take into account redshift distortions and galaxy bias with respect to matter. Again this can be done using $N$-body simulations, where dark matter halos are populated using a halo model.

Models of galaxy correlation with BAOs are constructed using the linear matter correlation function with non-zero baryon fraction $\Omega_b>0$, and further applying the different corrections. On the other hand, models of galaxy correlation without BAOs can be obtained by setting the baryon fraction to zero $\Omega_b=0$. One can also construct no-BAO models with $\Omega_b>0$, by using only the non-oscillatory part of the power spectrum to remove the effect of BAOs (e.g. the no-wiggles fit of \cite{Eis98}). In this case, BAOs are artificially erased and the models are non-physical. Yet they enable to test the existence of BAOs independently of other baryonic effects.

\subsection{BAO detection by hypothesis testing}
\label{hypothesis_section}
Let us define the two different hypotheses
\begin{eqnarray*}
\mathcal{H}_0 & : &   \mbox{no-BAO hypothesis} \\
\mathcal{H}_1 & : &   \mbox{BAO hypothesis} 
\end{eqnarray*}

BAO detection is equivalent to this problem of hypothesis testing. The common procedure is to design a \textit{test statistic} to assess the truth of the null hypothesis $\mathcal{H}_0$. From the test statistic obtained with the measurement, one computes a $p$-value and a significance. If the measurement if found to be more unlikely than a given threshold under $\mathcal{H}_0$, one rejects $\mathcal{H}_0$ and accepts $\mathcal{H}_1$ (see section \ref{chi2_limitations}).

We focus on the case where the data measurement is the correlation function estimated in different bins $\hat{\xi}=(\hat{\xi}_i)_{1\le i \le n}$. Such a binning is always present for the measurement, and thus for the model correlations in the hypotheses $\mathcal{H}_0$ and $\mathcal{H}_1$. As a slight abuse of language, we will use the terms estimated correlation function and model correlation functions for designating their binned versions. 

The hypotheses $\mathcal{H}_0$ and $\mathcal{H}_1$ are based on BAO and no-BAO models of correlation function, as the ones presented section \ref{correl_model}. The hypotheses also include the noise of the measurement, i.e. the covariance matrix $C=(C_{i,j})_{1\le i,j \le n}$ of $\hat{\xi}$. 

We will see in section \ref{wavelet_section} that wavelet methods are mainly sensitive to the BAO feature in the correlation function, and not on the global shape of the model. 

On the other hand, usual detection methods (e.g. the $\chi^2$ method) are based on full modeling of the correlation function. In this case, BAO and no-BAO models of correlation function $\xi_{BAO,\theta}$ and $\xi_{noBAO,\theta}$ are parameterized by $\theta$ to account for variations of cosmological parameters. The hypotheses are 
\begin{eqnarray*}
\mathcal{H}_0 & : &  \exists \, \theta \in \Theta \,\, \mbox{s.t.} \,\, \hat{\xi} \thicksim \mathcal{N}\left( \xi_{noBAO,\theta} ,C_{noBAO,\theta} \right) \\
\mathcal{H}_1 & : &  \exists \, \theta \in \Theta \,\, \mbox{s.t.} \,\, \hat{\xi} \thicksim \mathcal{N}\left( \xi_{BAO,\theta} ,C_{BAO,\theta} \right) 
\end{eqnarray*}

Most methods work with these hypotheses, where $\hat{\xi}$ is Gaussian. Ideally the hypotheses are sampled by $N$-body simulations for each model, which does not force $\hat{\xi}$ to be Gaussian. Yet we will see in section \ref{verif_gaussian_simu} that the Gaussian approximation works very well for our lognormal simulations.

The classical $\chi^2$ method used for BAO detection that we present in section \ref{chi2_section} simplifies the hypotheses by assuming constant covariance matrices ( i.e. independent of the model). The reason is that it can be hard to evaluate the covariance matrix for all tested models. In a modified version of the $\chi^2$ method that we propose in section \ref{modified_chi2_section} this simplification is not imposed.

The distribution of $\hat{\xi}$ is entirely specified for each hypothesis when fixing $\theta$, which is not the case when allowing $\theta$ to vary. In the first case the hypotheses are said to be simple, whereas in the second case they are said to be composite\footnote{In statistics an hypothesis is said to be simple when the distribution of the random variable is completely specified. It is said to be composite when the distribution is not completely specified. For example a union of simple hypotheses gives a composite hypothesis.}. 

BAO detection makes more sense when allowing variations of $\theta$, since it takes into account  uncertainties in cosmological parameters. However we will see in section \ref{chi2_section} with the classical $\chi^2$ method, that one has to be careful when testing composite hypotheses.

\section{Wavelet filtering methods}
\label{wavelet_section}
As explained in section \ref{correl_model}, BAOs manifest as a characteristic peak in the correlation function at the acoustic scale $r_s$. For detecting this feature new methods have recently emerged, based on wavelet analysis \citep{Xu10,Tia11,Arn11}. Wavelet transforms are widely used in many areas, especially in image analysis \citep{Mal08,Sta10}. They are specially suited for the analysis of data at different scales, and identification of characteristic patterns or structures. 

Here the characteristic structure is the BAO feature in the correlation, with uncertainty in its scale and shape. Uncertainty in the scale comes from a wrong fiducial cosmology to convert redshifts into distances, and a weak dependence of $r_s$ on cosmological parameters. Uncertainty in the shape is due to non-linear evolution, redshift distortions which are subject to modeling errors (see section \ref{correl_model}).

\begin{figure}  
\plotone{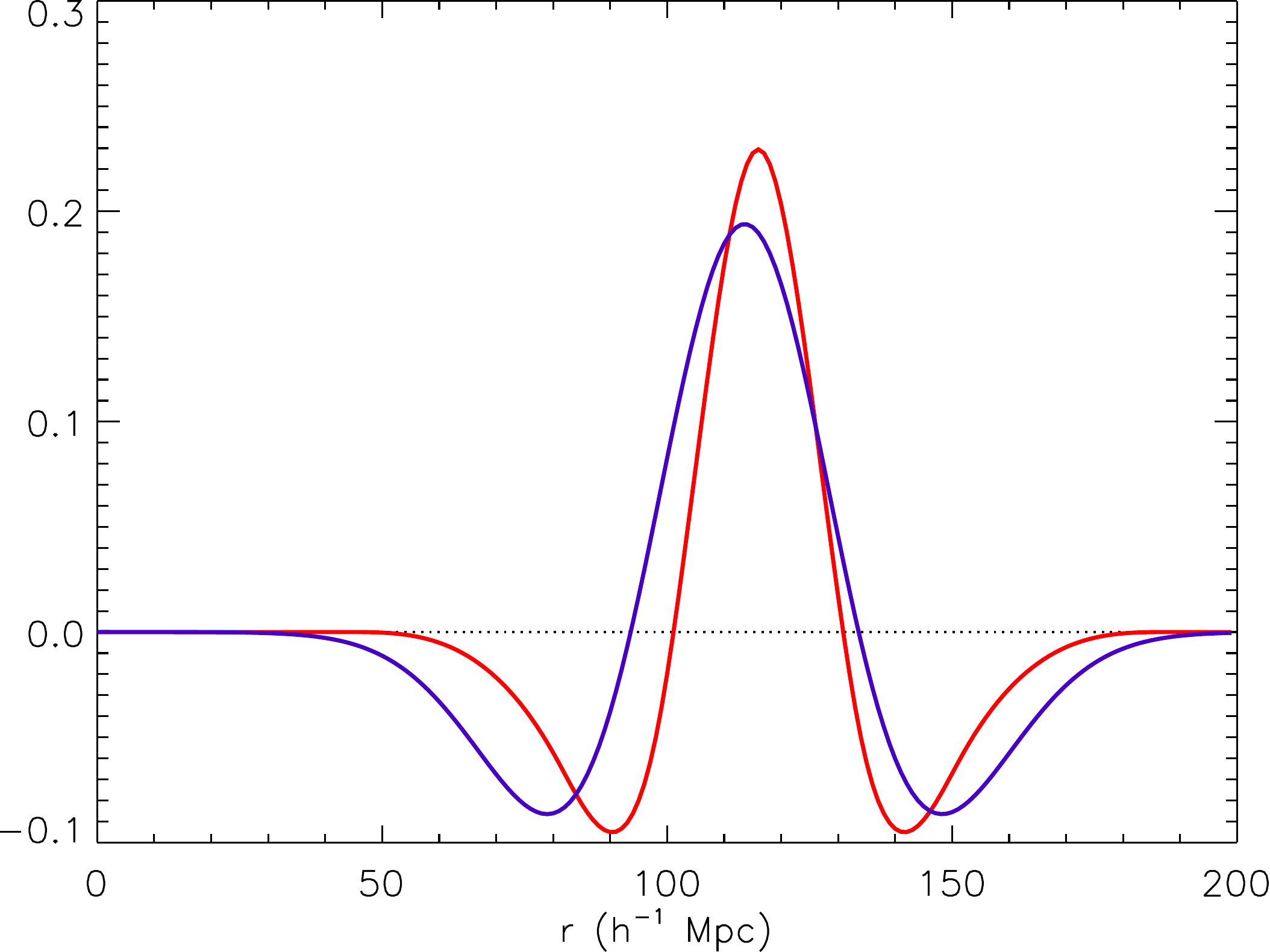} 
\caption{Different analyzing wavelets used for BAO detection. We show the mexican hat with parameters $R=113.6 \, h^{-1} \mbox{Mpc}$, $s=20 \, h^{-1} \mbox{Mpc}$ (blue) and the BAOlet with parameters $R=116\, h^{-1} \mbox{Mpc}$, $s=36 \, h^{-1} \mbox{Mpc}$ (red). These parameters correspond to the maximum responses $Z^{max}_w$ in the respective studies \cite{Tia11} and \cite{Arn11}.} 
\label{wavelet} 
\end{figure}

Our focus here is BAO detection, so we present two different methods that have been developed for this purpose. In these methods, a wavelet $w=(w(r_i))_{1\leq i\leq n}$ acts as a peak finder to detect excess in the measured correlation $(\hat{\xi}(r_i))_{1\leq i\leq n}$. The wavelet $w(R,s)$ is parametrized by two parameters $R$ and $s$, linked respectively to the scale and width of the peak. A simple way to do it is to consider an original peak-finding wavelet $w_0$ and rescale it as $w_i(R,s)=w_0\left(\frac{r_i-R}{s}\right)$ (see figure \ref{wavelet}).

One obtains a filtered signal $S_w(R,s)$ for every pair $(R,s)$
\begin{equation}
S_w(R,s)  = \left \langle w(R,s), \hat{\xi} \right \rangle = \sum^n_{i=1} w_i(R,s) \hat{\xi}(r_i) 
\end{equation}

The next step is to divide $S_w(R,s)$ by its noise $\sigma_w(R,s)$ under $\mathcal{H}_0$ to obtain $Z$-scores $Z_w(R,s)$. The parameters $(R_{max},s_{max})$ giving the maximum $Z$-score can be used to estimate the BAO scale and width.
\begin{eqnarray}
Z_w(R,s) & = &\frac{S_w(R,s)}{\sigma_w(R,s)} \\
Z^{max}_w & = & Z_w(R_{max},s_{max})
\end{eqnarray}

The two hypotheses are roughly that the maximum response $Z^{max}_w$ is negligible under $\mathcal{H}_0$ (no peak is found), and that there is a non-negligible signal under $\mathcal{H}_1$ (a peak is found). To reject $\mathcal{H}_0$ one performs simulations without BAOs, and computes how rarely a value of $Z^{max}_w$ as high as in the data is observed. This gives a $p$-value and thus the significance of the detection.

The major advantage of using wavelets is that they are mainly sensitive to the BAO feature, and not to smooth changes in the correlation function. Furthermore, scale variations of the analyzing wavelets allows dilations of the correlation function (to probe different cosmologies), and variations of its width allows for variations in the shape of the BAO peak. The consequence is that the wavelet response is mainly related to the existence of a BAO peak, and mildly dependent on the whole modeling of the correlation function. In other words, wavelet methods are robust to small modeling errors in the correlation function.

The price to pay is that they are outperformed by some model-dependent methods, when there are no modeling errors in $\mathcal{H}_0$ and $\mathcal{H}_1$. As we will see in section \ref{modified_chi2_section}, the $\Delta \chi^2$ (respectively $\Delta l$)  statistic can be seen as a generalized likelihood ratio between $\mathcal{H}_0$ and $\mathcal{H}_1$ in the case of constant (respectively varying) covariance matrix. The interest of the likelihood ratio is that it is optimal in the Neyman-Pearson sense for simple hypotheses (see section \ref{chi2_limitations} and appendix \ref{optimality_lratio}). For composite hypotheses there is no such notion of optimality. But we also find in section \ref{test_simu} that the generalized likelihood ratio $\Delta l$ gives better results than $\Delta \chi^2$ in the case of varying covariance matrix. Since wavelet statistics are not directly linked to the likelihood ratio, we expect less good results than with a generalized likelihood ratio.

A wavelet detection method was used in \cite{Tia11} on the SDSS Main sample. An angular average of the 3D anisotropic correlation function is computed, by applying a flat weighting in all directions, instead of the angular average on the sphere. The justification is that the BAO feature is sharpened in the line-of-sight direction, and this type of weighting is giving more importance to it. The correlation is further analyzed with a mexican hat wavelet (see figure \ref{wavelet}). Using Gaussian simulations with a no-wiggles power spectrum, they find that only $0.2 \%$ of the simulations have the statistic $Z^{max}_w$ at the same level as the data, which means a $3.1 \sigma$ detection.

In \cite{Arn11} two different galaxy samples of the SDSS DR7 are used, to compute the mean density profile of the Main sample around LRGs. This signal is analyzed with a 3D wavelet called BAOlet (see figure \ref{wavelet}). Because the wavelet is isotropic, it is equivalent to applying a 1D wavelet transform on the cross-correlation LRG-Main. Simulations of $\mathcal{H}_0$ are made by replacing LRGs by random centers to show that LRGs are located at special positions. Using this hypothesis, a signal as high as $Z^{max}_w$  is found with a probability $p=4.10^{-5}$ in simulations, corresponding to a $4.1 \sigma$ detection. Again this cannot be compared to other existing methods, because the tested hypothesis $\mathcal{H}_0$ is very different. \\

\section{$\chi^2$ method}
\label{chi2_section}

The $\chi^2$ method is the classical method used for BAO detection and can deal with the general case of varying cosmological parameters. Unlike wavelets methods presented in section \ref{wavelet_section} it is fully model-dependent, so it is mainly useful when all effects in the correlation function are well understood.

The $\chi^2$ method is also designed for hypotheses where the measurement $\hat{\xi}$ is Gaussian. In the rest of the paper we will only consider such hypotheses. In section \ref{verif_gaussian_simu}, we will see using simulations that the Gaussian approximation is well justified.

\subsection{The $\chi^2$ statistic}
For a measured correlation function $\hat{\xi} \thicksim \mathcal{N}\left(\xi_m,C \right)$ the $\chi^2$ statistic writes
 \begin{eqnarray}
 \label{chi2_def}
 \chi^2 & = &\left \langle \hat{\xi}-\xi_m, C^{-1} (\hat{\xi}-\xi_m) \right \rangle \\
  & = & \sum_{1\leq i,j\leq n}  \left[ \hat{\xi}(r_i)-\xi_m(r_i) \right] C^{-1}_{i,j}  \left[ \hat{\xi}(r_j)-\xi_m(r_j) \right]
 \end{eqnarray}
 
Supposing that the model is correct (i.e. $\hat{\xi} \thicksim \mathcal{N}\left(\xi_m,C \right)$), the $\chi^2$ statistic follows a $\chi_n^2$ distribution, i.e. a  chi-square distribution with $n$ degrees of freedom. The $\chi_n^2$ distribution can be interpreted as the one followed by the sum of the squares of $n$ independent standard normal variables. If $X_1$, \dots, $X_n$ are $n$ such i.i.d. Gaussian variables then $\sum^n_{i=1} X^2_i$ follows a $\chi^2_n$ distribution.

\subsection{$\chi^2$ method for BAO detection}
\label{chi2_detection}
Let us show how the BAO detection is usually performed. We note $\theta=(\theta_1, \dots, \theta_k)\in \Theta $ the dependence parameters for the model correlation functions with and without BAOs, $\xi_{BAO,\theta}$ and $\xi_{noBAO,\theta}$. The hypotheses are given by
\begin{eqnarray*}
\mathcal{H}_0 & : &  \exists \, \theta \in \Theta \,\, \mbox{s.t.} \,\, \hat{\xi} \thicksim \mathcal{N} \left( \xi_{noBAO,\theta}, C \right) \\
\mathcal{H}_1 & : &  \exists \, \theta \in \Theta \,\, \mbox{s.t.} \,\, \hat{\xi} \thicksim \mathcal{N} \left( \xi_{BAO,\theta}, C \right) 
\end{eqnarray*}

As mentioned in section \ref{hypothesis_section}, the $\chi^2$ method tests hypotheses with a constant covariance matrix $C$. The parameters $\theta$ are not directly cosmological parameters but are linked to them. For example in \cite{Eis05}, they are given by a dilation parameter $\alpha$ (to account for a wrong fiducial cosmology to convert redshifts into distances), an amplitude parameter $b^2$ (to account for galaxy bias, redshift distortions, and power spectrum normalization $\sigma_8$), and the parameter $\Omega_m h^2$ (determining the horizon scale at matter-radiation equality, the amplitude of the BAO peak, and more moderately the position of the peak). Other parameters also have an impact on the expected correlation function ($\Omega_b h^2$ and the spectral index $n$) but they are well constrained by CMB data and can be fixed as a good approximation.

The $\chi^2$ has a dependence on the parameters $\theta$
\begin{eqnarray}
\chi_{BAO,\theta}^2 &= & \left \langle \hat{\xi}-\xi_{BAO,\theta}, C^{-1} (\hat{\xi}-\xi_{BAO,\theta}) \right \rangle\\
\chi_{noBAO,\theta}^2 &= &  \left \langle \hat{\xi}-\xi_{noBAO,\theta}, C^{-1} (\hat{\xi}-\xi_{noBAO,\theta}) \right \rangle
\end{eqnarray}

For each class of models one can look at the $\chi^2$ best-fits, $\min_{\theta} \chi^2_{noBAO,\theta}$ and $\min_{\theta} \chi^2_{BAO,\theta}$. It is a widely used result that the best-fit $\chi^2$ value follows a $\chi_{n-k}^2$ distribution, assuming that the true model is inside the fitting class. So the number of degrees of freedom decreases by the number of parameters in the fit. We stress that this result (see appendix \ref{best_fit}) is only rigorously valid when the space of model correlations is affine. We recall that the measurement vector $\hat{\xi}$ and the model correlations $\xi_m(\theta)$ are $n$-dimensional binned versions of their continuous counterparts. So the set of all model correlations $\left(\xi_m(\theta) \right)_{\theta \in \Theta}$ constitutes a subspace of $\mathbb{R}^n$, which needs to be a $k$-dimensional affine space for the previous result to hold (see appendix \ref{best_fit}).

This result can be used on $\min_{\theta} \chi^2_{BAO,\theta}$ to verify that data are compatible with $\mathcal{H}_1$. More precisely it can be tested whether $\min_{\theta} \chi^2_{BAO,\theta}$ is compatible with its distribution when the true model is in $\mathcal{H}_1$
\begin{equation}
\min_{\theta} \chi^2_{BAO,\theta} \thicksim \chi^2_{n-k}
\end{equation}

For the rejection of $\mathcal{H}_0$, the usual procedure is more complex. We add an artificial parameter in the fit, which accounts for the presence of BAOs in the model correlation. For example this parameter $\beta$ can be a weighting of $\xi_{BAO,\theta}$ and $\xi_{noBAO,\theta}$ in the model correlation function.
\begin{equation}
\xi_{\beta,\theta} = \beta \, \xi_{BAO,\theta} + (1-\beta) \, \xi_{noBAO,\theta} 
\end{equation}

Under $\mathcal{H}_0$, the expected correlation function is of the form $\xi_{noBAO,\theta}$ for the true parameters $\theta=\theta_0$, but it is also of the form $\xi_{\beta,\theta}$ with $\beta=0$ and $\theta=\theta_0$. Thus the best-fit $\chi^2$ value in the no-BAO class follows a $\chi_{n-k}^2$ distribution, and the best-fit $\chi^2$ value in the extended class follows a $\chi_{n-(k+1)}^2$ distribution (since $\beta$ is an additional parameter). 

We are in the case of two nested classes of models, which both contain the true model. In this case, the difference of the best-fit values  follows a chi-square distribution with number of freedom equal to the difference of parameters between the two classes. Again this result is not rigorously true in the general case, but only when the two spaces of model correlations are affine (see appendix \ref{nested_best_fit}). 

Here there is only one additional parameter, $\beta$, in the extended class, thus the difference of the best-fit values follows a $\chi_1^2$ distribution. 
\begin{equation}
\Delta \chi^2_{global}=\min_{\theta} \chi_{noBAO,\theta}^2-\min_{\beta,\theta} \chi_{\beta,\theta}^2 \thicksim \chi_1^2
\label{deltachi2_global}
\end{equation}
This accounts for the fact that fitting an additional parameter, which is not required by the true model, only moderately decreases the best-fit value. To reject $\mathcal{H}_0$, one can look at this difference $\Delta \chi^2_{global}$ and compute how unlikely it is under $\mathcal{H}_0$ (i.e for a $\chi_1^2$ distribution). 

In practice, the best-fit in the whole extended class is replaced by the best-fit in the BAO class (i.e. restricting to $\beta=1$)
\begin{equation}
\Delta \chi^2=\min_{\theta}  \chi^2_{noBAO,\theta}-\min_{\theta} \chi^2_{BAO,\theta}
\end{equation}
This difference $\Delta \chi^2$ is necessarily less than $\Delta \chi^2_{global}$ of equation (\ref{deltachi2_global}), which follows a $\chi_1^2$ distribution under $\mathcal{H}_0$. 

Thus for a realization with an arbitrary value $\Delta \chi^2=x$, we have 
\begin{small}
\begin{equation}
P(\Delta \chi^2 > x \, | \, \mathcal{H}_0) \leq P(\Delta \chi^2_{global} > x \, | \, \mathcal{H}_0)
\label{inequality}
\end{equation}
\end{small}
A $\chi_1^2$ distribution is simply the distribution followed by the square of a standard normal variable.  Noting $\Phi$ the cumulative distribution function of a standard Gaussian variable, we get for $x \geq 0$ 
\begin{small}
\begin{equation}
P(\Delta \chi^2 > x \, | \, \mathcal{H}_0) \leq P(\Delta \chi^2_{global}  > x \, | \, \mathcal{H}_0) = 2\Phi(-\sqrt{x})
\end{equation}
\end{small}
which corresponds to a number of $\sigma$ equal to $\sqrt{x}$ for a normal distribution. Thus, when $\Delta \chi^2 \geq 0$, one can evaluate the significance of the BAO detection as $\sqrt{\Delta \chi^2} . \sigma$. A significance is originally given in terms of a $p$-value, i.e. the probability of obtaining the measurement value under $\mathcal{H}_0$. When given as a number of $\sigma$, it is simply the corresponding number of standard deviation to the mean for a Gaussian variable.

In \cite{Eis05} the difference of chi-square equals $\Delta \chi^2=11.7$, corresponding to a $3.4 \sigma$ detection using this method. In \cite{Per10} it is equal to $\Delta \chi^2=13.1$, corresponding to a $3.6 \sigma$ detection. 

Because of the inequality in equation (\ref{inequality}) the method may seem conservative. However because the assumptions of the method are not verified, we will see in section \ref{classical_chi2_simu} that the method actually overestimates the significance.

\subsection{Limitations of the $\chi^2$ method}
\label{chi2_limitations}

The Neyman-Pearson lemma states that, when performing a hypothesis test between $\mathcal{H}_0$ and $\mathcal{H}_1$, the most powerful tests are based on the likelihood ratio 
\begin{equation}
\Lambda(\hat{\xi})=\frac{\mathcal{L}_{\mathcal{H}_0}(\hat{\xi})}{\mathcal{L}_{\mathcal{H}_1}(\hat{\xi})}
\end{equation}

More precisely, the most powerful test of significance $\alpha$ is 
\begin{itemize}
\item if $\Lambda(\hat{\xi}) \leq \eta$ then accept $\mathcal{H}_1$ (i.e. reject $\mathcal{H}_0$)
\item if $\Lambda(\hat{\xi}) > \eta$ then accept $\mathcal{H}_0$ (i.e. reject $\mathcal{H}_1$)\\
\end{itemize}

with $\alpha$ the probability of rejecting $\mathcal{H}_0$ if it is true (type I error)
\begin{equation}
\alpha=P\left(\Lambda(\hat{\xi}) \leq \eta \, | \, \mathcal{H}_0 \right)
\end{equation}

The power of the test is defined as the probability of accepting $\mathcal{H}_1$ if it is true. It is equal to $1-\beta$ where $\beta$ is the probability of type II error, i.e. the probability of not accepting $\mathcal{H}_1$ if it is true. 
\begin{equation}
\beta=P\left(\Lambda(\hat{\xi}) > \eta \, | \, \mathcal{H}_1 \right)
\end{equation}

In practice, such tests with given thresholds are not really used, and it is more common to cite the significance level for the realization value. For a realization with an arbitrary value $\Lambda(\hat{\xi})=x$, the significance (given as a $p$-value) is
\begin{equation}
\alpha(x)=P\left(\Lambda(\hat{\xi}) \leq x \, | \, \mathcal{H}_0 \right)
\label{pvalue_test}
\end{equation}

As we show in appendix \ref{optimality_lratio}, the Neyman-Pearson lemma implies that the expected significance under $\mathcal{H}_1$ obtained with $\Lambda(\hat{\xi})$ is better than with any other statistic. More precisely, under $\mathcal{H}_1$, the expected $p$-value of equation (\ref{pvalue_test}) is smaller, and the expected number of $\sigma$ is larger for $\Lambda(\hat{\xi})$ than for any other statistic.

Note that the statistic $\Lambda(\hat{\xi})$ is optimal in this sense, but we need to know its distribution under $\mathcal{H}_0$ (to compute the significance $\alpha(x)$ corresponding to a realization value $x$). Moreover in the case of composite hypotheses, the distribution of $\Lambda(\hat{\xi})$ is not well-defined under $\mathcal{H}_0$ (see section \ref{hypothesis_section}) so the significance is also not well-defined. The advantage of $\Delta \chi^2_{global}$ is that its distribution is identical for every model in $\mathcal{H}_0$ (a $\chi^2_1$ distribution). In this case one is able to give a significance even with composite hypotheses. Yet this result is subject to a regularity assumption, that spaces of model correlation functions are affine. Because it is not verified, we will see with simulations in section \ref{classical_chi2_simu} that the distribution of $\Delta \chi^2_{global}$ is quite different from a $\chi^2_1$ and that the $\chi^2$ method only gives a rough estimate of the significance.

The estimate can be even more wrong when considering more realistic hypotheses, where covariance matrices depend on the model 
\begin{eqnarray*}
\mathcal{H}_0 & : &  \exists \, \theta \in \Theta \,\, \mbox{s.t.} \,\, \hat{\xi} \thicksim \mathcal{N} \left( \xi_{noBAO,\theta}, C_{noBAO,\theta} \right) \\
\mathcal{H}_1 & : &  \exists \, \theta \in \Theta \,\, \mbox{s.t.} \,\, \hat{\xi} \thicksim \mathcal{N} \left( \xi_{BAO,\theta}, C_{BAO,\theta} \right) 
\end{eqnarray*}

Let us consider a common case where the parameters $\theta$ are given by $\theta=(\Omega_m h^2,\alpha,b)$ as introduced in section \ref{chi2_detection}. For illustrative purposes, we only take into account the dependence in the bias $b$ of the covariance matrix. It gives a multiplicative factor $b^2$ in the expected correlation function, and at first approximation (up to shot noise) a multiplicative factor $b^4$ in the covariance matrix. So the covariance matrices are given by 
\begin{equation}
C_{BAO,\theta}=C_{noBAO,\theta} \propto b^4 C
\end{equation}

In the classical $\chi^2$ method, $\chi^2_{BAO,\theta}$ and $\chi^2_{noBAO,\theta}$ are computed with a constant covariance matrix $C$. So for a realization $A \hat{\xi}$ with $A\geq 0$, we get the BAO best-fit

\begin{align}
& \min_\theta \chi^2_{BAO,\theta}(A\hat{\xi}) \nonumber \\
& \,\,=  \min_\theta \left \langle A\hat{\xi}-\xi_{BAO,\theta}, C^{-1} (A\hat{\xi}-\xi_{BAO,\theta}) \right \rangle \nonumber  \\
& \,\,=  A^2\min_\theta \left \langle \hat{\xi}-\frac{1}{A}\xi_{BAO,\theta}, C^{-1} \left(\hat{\xi}- \frac{1}{A} \xi_{BAO,\theta}\right) \right \rangle \nonumber  \\
& \,\,=  A^2 \min_\theta \chi^2_{BAO,\theta}(\hat{\xi})
\end{align}

The last equality comes from the role of $b$, which enables any positive multiplication of the model. The same reasoning can be applied to $\min_\theta \chi^2_{noBAO,\theta}$ which gets multiplied by $A^2$, and thus the statistic $\Delta \chi^2$ also gets multiplied by $A^2$. 

Given the hypotheses with varying covariance matrix, $\mathcal{H}_0$ realizations with $\theta=(\Omega_m h^2,\alpha,b_1)$ can be obtained from $\mathcal{H}_0$ realizations with $\theta=(\Omega_m h^2,\alpha,b_0)$  by a multiplicative factor $(b_1/b_0)^2$. As a result, the distribution of $\Delta \chi^2$ gets dilated by a factor $(b_1/b_0)^4$. This creates very large differences between the different models in $\mathcal{H}_0$, so the classical $\chi^2$ method provides very bad estimates of the significance. The conclusion is that the classical $\chi^2$ method cannot be used in the case of varying covariance matrix.

\section{Modified $\chi^2$ method}
\label{modified_chi2_section}
In this section we propose two modifications to the $\chi^2$ methods to overcome its limitations. A first modification enables to obtain the correct significance in all cases. So unlike the classical $\chi^2$ method, our modified method can be applied to hypotheses with varying covariance matrices
\begin{eqnarray*}
\mathcal{H}_0 & : &  \exists \, \theta \in \Theta \,\, \mbox{s.t.} \,\, \hat{\xi} \thicksim \mathcal{N} \left( \xi_{noBAO,\theta}, C_{noBAO,\theta} \right) \\
\mathcal{H}_1 & : &  \exists \, \theta \in \Theta \,\, \mbox{s.t.} \,\, \hat{\xi} \thicksim \mathcal{N} \left( \xi_{BAO,\theta}, C_{BAO,\theta} \right) 
\end{eqnarray*}

The way to rigorously compute the $p$-value for a measurement $\Delta \chi^2=x$ is to consider the "worst-case" $\mathcal{H}_0$ model
\begin{equation}
p(x)=\max_{\theta \in \Theta} P(\Delta \chi^2 \ge x \, | \, \mathcal{H}_0, \theta)
\label{pvalue_modified_chi2}
\end{equation} 

So for every model in $\mathcal{H}_0$, the $p$-value of the measurement $\Delta \chi^2=x$ is less than $p(x)$. When considering the significance $s(x)\sigma$ corresponding to this $p$-value, we get that every model in $\mathcal{H}_0$ is at least rejected at $s(x)\sigma$. Note that this is the best significance we can get to reject all $\mathcal{H}_0$ models simultaneously.

This way of obtaining the significance does not rely on any assumption, unlike in the classical $\chi^2$ method. However it requires more work to compute the distribution of $\Delta \chi^2$ under every $\mathcal{H}_0$ model. We will see how precisely this can be achieved in section \ref{modified_chi2_simu}, when applying the procedure on simulations.

The second modification we propose is on the statistic itself. As we saw in section \ref{chi2_limitations} the optimal statistic to test simple hypotheses $\mathcal{H}_0$ and $\mathcal{H}_1$ is the likelihood ratio. However when working with composite hypotheses, likelihoods are not well defined. Indeed they can be defined for any model in $\mathcal{H}_0$ or $\mathcal{H}_1$ but not for $\mathcal{H}_0$ and $\mathcal{H}_1$ themselves. The idea of the $\Delta \chi^2$ statistic is that it can be thought as a generalized likelihood ratio between composite hypotheses. Indeed in the special case of a constant covariance matrix we have
\begin{small}
\begin{eqnarray}
\Delta \chi^2 & = & \min_\theta \chi^2_{noBAO,\theta}-\min_\theta \chi^2_{BAO,\theta} \nonumber \\
 		     & = & -2\left[\max_\theta \ln \left( \mathcal{L}_{noBAO,\theta} \right) - \max_\theta \ln \left( \mathcal{L}_{BAO,\theta}\right) \right] \nonumber \\
		     & = & -2\ln \left[\frac{\max_\theta \mathcal{L}_{noBAO,\theta}}{\max_\theta \mathcal{L}_{BAO,\theta}} \right]
		\label{Dchi2L}
\end{eqnarray} 
\end{small}

where we used the relation between $\chi^2$ and the likelihood of equation (\ref{chi2L}). This is only valid for a constant covariance matrix. To extend this idea in the case of varying covariance matrices, we simply consider the statistic \\
\begin{equation}
\Delta l  = -2\left[\max_\theta \ln \left( \mathcal{L}_{noBAO,\theta} \right) - \max_\theta \ln \left( \mathcal{L}_{BAO,\theta}\right) \right]\\
\end{equation}
We use the notation $\Delta l$ because it refers to a difference of log-likelihoods. It is a slight abuse of notation since it is not strictly speaking a difference of log-likelihoods. Unlike $\Delta \chi^2$, $\Delta l$ is still a generalized likelihood ratio for varying covariance matrices. So it should give better results in this case as we verify in section \ref{detect_varcov_simu}.

In the case of varying covariance matrices the likelihoods write 
\begin{eqnarray}
 \mathcal{L}_{BAO,\theta} & \propto & |C_{BAO,\theta}|^{-1/2} \, e^{ -\frac{\chi^2_{BAO,\theta} }{2}   } \\
 \mathcal{L}_{noBAO,\theta} & \propto & |C_{noBAO,\theta}|^{-1/2} \, e^{ -\frac{\chi^2_{noBAO,\theta} }{2}   } \\ \nonumber
\end{eqnarray}
where $\chi^2_{BAO,\theta}$ and $\chi^2_{noBAO,\theta}$ are computed by taking into account variations of the covariance matrix
\begin{small}
\begin{eqnarray}
\label{chi2_extended1}
\chi^2_{BAO,\theta}  & = & \left \langle \hat{\xi}-\xi_{BAO,\theta}, C_{BAO,\theta}^{-1} \, (\hat{\xi}-\xi_{BAO,\theta}) \right \rangle \\
\label{chi2_extended2}
\chi^2_{noBAO,\theta}  & = & \left \langle \hat{\xi}-\xi_{noBAO,\theta}, C_{noBAO,\theta}^{-1} \, (\hat{\xi}-\xi_{noBAO,\theta}) \right \rangle 
\end{eqnarray}
\end{small}

Let us write $l_{noBAO,\theta}$ and $l_{BAO,\theta}$ for $-2 \ln \left(\mathcal{L}_{noBAO,\theta} \right)$ and $-2 \ln \left(\mathcal{L}_{BAO,\theta} \right)$ we get

\begin{eqnarray}
 \label{delta_l}
\Delta l             & = & \min_\theta l_{noBAO,\theta} - \min_\theta l_{BAO,\theta}  \\
 \label{logL_extended1}
l_{BAO,\theta}  & = & \chi^2_{BAO,\theta} + \ln  |C_{BAO,\theta}| +cst \\
 \label{logL_extended2}
l_{noBAO,\theta}  & = & \chi^2_{noBAO,\theta} + \ln  |C_{noBAO,\theta}| +cst\\ \nonumber
\end{eqnarray}

with the same additive constant for $l_{BAO,\theta}$ and $l_{noBAO,\theta}$, which can be taken as 0. 

Note that $\Delta l$ is \textit{not} equivalent to $\Delta \chi^2$ even if $\chi^2_{BAO,\theta}$ and $\chi^2_{noBAO,\theta}$ are computed using equations (\ref{chi2_extended1}) and (\ref{chi2_extended2}). Indeed in the case of varying covariance matrices, one has to take into account variations of the matrix determinant as in equations (\ref{logL_extended1}) and (\ref{logL_extended2}).

We will refer to the $\Delta l$ method for the $\chi^2$ method modified as we suggested: replacing $\Delta \chi^2$ by $\Delta l$ \textit{and} modifying the procedure to obtain the correct significance.

\section{Cosmological parameters constraints}
\label{constraints_section}
Let us describe the second use of BAOs, which can help constrain cosmological parameters. Here the true cosmological model  is supposed to be in $\mathcal{H}_1$.

\subsection{Constraints with constant covariance matrix}
\label{constraints_constant_cov}
In the case of a constant covariance matrix the hypotheses $\mathcal{H}_1$ is given by
\begin{eqnarray*}
\mathcal{H}_1 & : &  \exists \, \theta \in \Theta \,\, \mbox{s.t.} \,\, \hat{\xi} \thicksim \mathcal{N} \left( \xi_{BAO,\theta}, C \right) 
\end{eqnarray*}

For a given model in $\mathcal{H}_1$, the measurement $\hat{\xi}$ is Gaussian and $\chi_{BAO,\theta}^2$ is equivalent to the log-likelihood
\begin{equation}
\label{chi2L}
\chi_{BAO,\theta}^2 = - n \ln\left(2\pi |C|\right) - 2 \ln \left( \mathcal{L}_{BAO,\theta}(\hat{\xi}) \right)
\end{equation}

To define the posterior probability $p(\theta \, | \,\hat{\xi})$ one needs a prior $p(\theta)$ on $\theta$
\begin{eqnarray}
p(\theta \, | \,\hat{\xi}) & \propto & p(\theta) \, \mathcal{L}_{BAO,\theta}(\hat{\xi}) \\
							& \propto & p(\theta) \, \exp\left( -\frac{1}{2} \chi^2_{BAO,\theta} \right)
\end{eqnarray}

To obtain constraints only coming from $\hat{\xi}$ one can assume a constant prior $p(\theta)$ 
\begin{equation}
p(\theta \, | \,\hat{\xi}) \propto  \mathcal{L}_{BAO,\theta}(\hat{\xi})
\end{equation}

Note that this choice is still arbitrary because it is linked to a given parameterization. Indeed a constant prior $p(\theta)$ can lead to a non-constant prior for a different parameterization.

To combine constraints from $\hat{\xi}$ with the ones from other independent experiments, one has to modify the prior. For example with CMB data the posterior is given by
\begin{eqnarray}
p(\theta \, | \, \mbox{CMB}, \hat{\xi}) & \propto & p(\theta, \mbox{CMB}, \hat{\xi} ) \nonumber  \\
 					& \propto & p(\theta, \mbox{CMB}) \, p(\hat{\xi} \, | \, \theta,\mbox{CMB}) \nonumber \\
					 & \propto & p(\theta \, | \, \mbox{CMB}) \,  \mathcal{L}_{BAO,\theta}(\hat{\xi})
\end{eqnarray}
where we used the independence of $\hat{\xi}$ and CMB measurement. Adding the CMB measurement is thus equivalent to using a prior $p(\theta)=p(\theta\, | \, \mbox{CMB})$.
 
Again we consider the parameters $\theta$ given by $\theta=(\Omega_m h^2,\alpha,b)$. The parameter $\alpha$ accounts for dilation of the correlation function and $b$ corresponds to a multiplicative factor $b^2$. The correlation function models are thus given by
\begin{equation}
\xi_{BAO,\theta}(r)  =  b^2 \, \xi_{BAO,\Omega_m h^2}(\alpha \, r) 
\end{equation}

$\alpha$ is linked to the dilation scale $D_V(z)$ at the mean redshift of the sample $z$ by the relation $\alpha=D_V(z)/D_{V,fid}(z)$, with $D_{V,fid}(z)$ the dilation scale for the fiducial cosmology used to construct the 3D data catalogue. The dilation scale expresses how distances dilate when modifying the fiducial cosmology \citep{Eis05}. It depends on the Hubble parameter (line-of-sight dilation) and the transverse comoving distance $D_M(z)$ (transverse dilation)
\begin{equation}
D_V(z)=\left[  D_M(z)^2 \, \frac{cz}{H(z)} \right]^{1/3}
\end{equation}

One is interested in constraining $\Omega_m h^2$ and $D_V(z)$. We consider a constant prior $p(\theta)$ to obtain constraints only from $\hat{\xi}$. The posterior $p(\Omega_m h^2,\alpha \, | \,  \hat{\xi})$ is obtained by marginalizing over the multiplicative factor $B=b^2$
\begin{small}
\begin{eqnarray}
p(\Omega_m h^2,\alpha \, | \,\hat{\xi}) & \propto & \int  \mathcal{L}_{BAO,(\Omega_m h^2,\alpha,B)} dB \\
 & \propto & \int  \exp\left(-\frac{1}{2} \chi^2_{BAO,(\Omega_m h^2,\alpha,B)} \right) dB 
\end{eqnarray}
\end{small}
The posterior of $\Omega_m h^2$ is obtained by marginalizing $p(\Omega_m h^2,\alpha \, | \,  \hat{\xi})$ over $\alpha$, and the posterior of $\alpha$ by marginalizing over $\Omega_m h^2$. For each parameter, the maximum in the posterior gives the parameter estimate and the standard deviation can give a $1\sigma$ interval.

\subsection{Constraints with varying covariance matrix}
In the case of varying covariance matrix the hypothesis $\mathcal{H}_1$ is given by
\begin{eqnarray*}
\mathcal{H}_1 & : &  \exists \, \theta \in \Theta \,\, \mbox{s.t.} \,\, \hat{\xi} \thicksim \mathcal{N} \left( \xi_{BAO,\theta}, C_{BAO,\theta} \right) 
\end{eqnarray*}

In this case one has to take into account the dependence of the covariance matrix on the model
\begin{eqnarray}
 \chi^2_{BAO,\theta}  & = & \left \langle \hat{\xi}-\xi_{BAO,\theta}, C_{BAO,\theta}^{-1} \, (\hat{\xi}-\xi_{BAO,\theta}) \right \rangle \\
 \label{correct_likelihood}
 \mathcal{L}_{BAO,\theta} & \propto & |C_{BAO,\theta}|^{-1/2} \, e^{ -\frac{\chi^2_{BAO,\theta} }{2}   }
\end{eqnarray}

Let us consider the simple dependence $C_{BAO,\theta} \propto b^4 C$ as an illustration again. In this case, the marginalization over $b^2$ gives a different result compared to the result with constant covariance. So the obtained posteriors of $\Omega_m h^2$ and $\alpha$ are also different. We will see with simulations in section \ref{constraints_varcov_simu} that this changes indeed the constraints.

\section{Tests on simulations}
\label{test_simu}

\subsection{Simulations}
We use the same procedure for generating lognormal simulations of the SDSS DR7 LRG sample as in \cite{Lab11} with only small differences in the input power spectrum. Because we consider a volume-limited LRG sample with only the northern contiguous region, the volume is approximately half of the the full LRG sample and the number of galaxies a third. In particular the expected detection significance is lower than for the full LRG sample. So our focus is not on the expected detection significance for current surveys, but rather on comparing the different methods.

We use a $\Lambda$CDM power spectrum given by the iCosmo software \citep{Ref11}, with parameters $h=0.7$, $\Omega_b=0.045$, $\Omega_m=0.27$, $\Omega_\Lambda=0.73$, $n_s=1.0$, $\sigma_8=0.8$, and taken at redshift $z=0.3$. The transfer function is the form with wiggles of \cite{Eis98} and the non-linear correction to the power spectrum is obtained using the fitting formula of \cite{Smi03}. We model the non-linear degradation of the acoustic peak by multiplying the part of the power spectrum containing the oscillations by the function $\exp(-a^2 k^2/2)$ with $a=7 h^{-1}$Mpc (i.e. smoothing the oscillations in the correlation with a Gaussian of width $a$). This is found to be a good approximation in \cite{Eis07}. It consists more precisely in constructing the power spectrum using the forms with and without wiggles
\begin{small}
\begin{equation}
P(k)=P_{no\,wig}(k)+\exp(-a^2 k^2/2) [ P_{wig}(k)-P_{no\,wig}(k)]
\end{equation}
\end{small}
We further apply a constant bias $b^2$ with $b=2.5$ to this power spectrum so that the corresponding correlation function matches the one estimated on the SDSS LRG sample. Our simulations do not take into account the scale-dependence of the galaxy bias. However this is not a problem here since we only use simulations, and do not analyze real data.

We estimated the correlation function on 2000 independent simulations using each time the Landy-Szalay estimator and 100.000 random points. With this procedure Landy-Szalay has been shown to be the estimator with minimum variance, and to be nearly unbiased (see \cite{Lab11}). 

We estimated the covariance matrix by the empirical covariance matrix of the measured correlation function in the simulations. We use bins of size $10 \, h^{-1}$Mpc and perform the analysis in the range $20$ to $200 \, h^{-1}$Mpc. In this way we obtain $n=18$ bins, corresponding to 171 free parameters in the covariance matrix. This is much smaller than the number of simulations so that the empirical covariance matrix gives a good estimate of the true covariance matrix (see e.g. \cite{Pop08}). Another reason for not using a strong binning, is that most methods use the inverse of the covariance matrix. Since the bins are very correlated at small separation, the inverse matrix is very oscillating for too small binning, and the result becomes non robust to small modeling errors.

We find a small statistical bias in the correlation function estimation, due to the limited resolution of the lognormal simulations or the integral constraint (see \cite{Lab11}). This bias is negligible in absolute value ($\approx 5 \times 10^{-4}$), but when multiplied by $C^{-1}$ as in $Z_w$, $\chi^2_{BAO,\theta}$ or $\chi^2_{noBAO,\theta}$ it can slightly affect the result. Here the lognormal simulations will only be used for computing the covariance matrix and verifying the Gaussianity of $\hat{\xi}$. However when working with real data, one should verify that the integral constraint is not biasing results.

\subsection{Models}
Unless otherwise stated, we use for the BAO and no-BAO models in $\mathcal{H}_1$ and $\mathcal{H}_0$ the transfer function of \cite{Eis98} with respectively the form with wiggles and without wiggles. At some points we also quote the results obtained with no-BAO models constructed with zero baryon. 

The procedure for generating the power spectrums is the same as for the lognormal simulations. Obviously the non-linear degradation of the BAO peak is only performed for BAO models, by smoothing with a kernel of size $a$ in comoving coordinates.

The respective correlation functions $\xi_{BAO,\theta}$ and $\xi_{noBAO,\theta}$ are obtained from the power spectrums by inverse Fourier transform in 3 dimensions, i.e. by a Hankel transform in the isotropic case. The power spectrum has bins with exponential sizes in $k$ (i.e. the $\ln (k_i)$'s are spaced linearly) since it is smooth in that space. For doing the Hankel transform with this spacing we use the FFTLog code\footnote{http://casa.colorado.edu/$\sim$ajsh/FFTLog/}. The correlation is finally binned equivalently as when it is estimated by pair counting, i.e. for a bin $[r_i -dr/2,r_i +dr/2]$
\begin{equation}
\xi(r_i)= \frac{\int_{r_i -dr/2}^{r_i +dr/2}  \xi(r) \, r^2 \,  dr } {\int_{r_i-dr/2}^{r_i+dr/2}  \, r^2 \,  dr }
\end{equation}

We fix the parameters $\Omega_b h^2=0.0315$, $n_s=1.0$ and $\sigma_8=0.8$ as in the lognormal simulations, and choose the same parametrization as before $\theta=(\Omega_m h^2, \alpha, b)$ for the $\Lambda\mbox{CDM}$ correlation functions
\begin{eqnarray*}
\xi_{BAO,\theta}(r) & = & b^2 \, \xi_{BAO,\Omega_m h^2}(\alpha \, r) \\
\xi_{noBAO,\theta}(r) & = & b^2 \, \xi_{noBAO,\Omega_m h^2}(\alpha \, r) 
\end{eqnarray*}

To obtain these functions we vary $\Omega_m h^2$, adjust the value $\Omega_\Lambda=1-\Omega_m $, and perform the dilation in $\alpha$ of the correlation function obtained in comoving coordinates. There are different choices for varying $\Omega_m h^2$, and we choose to keep $h=0.7$ constant and vary $\Omega_m$. Another choice would be to keep $\Omega_m=0.27$ and vary $h$. These choices lead to different amplitudes of the correlation due to different growth factors at redshift $z=0.3$. We verified that our results are only slightly affected by this choice.

Finally the functions are rescaled by a factor $b^2$ for modeling the galaxy bias. We have to take into account that no-BAO models with zero baryon have different amplitude than the other models for the range of scales considered (for the same $\sigma_8$). Therefore we rescale them by an amplitude factor of $1.29$, which is found to minimize the distance between the no-BAO and BAO models for the parameter values of the lognormal simulations 
\begin{equation}
\left \langle \xi_{BAO}-\xi_{noBAO}, C^{-1} (\xi_{BAO}-\xi_{noBAO}) \right \rangle
\end{equation}

\begin{figure}  
\plotone{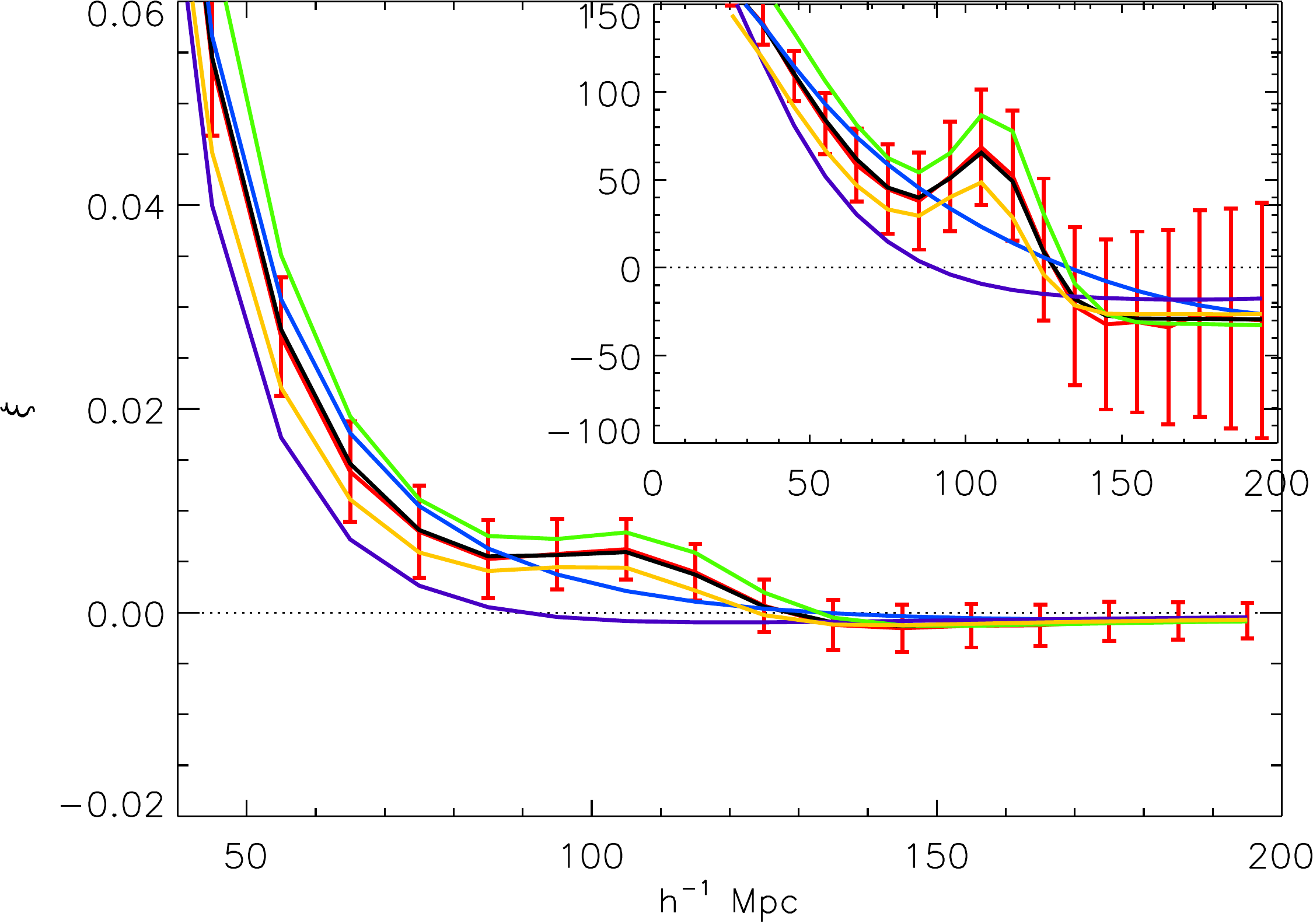} 
 \caption{Correlation function $\xi(r)$ as observed in the fiducial cosmology. In inset we plot $r^2 \xi(r)$ for a better visualization. For the lognormal simulation parameters $(\Omega_m h^2,\alpha,b)=(0.1323,1,2.5)$, we plot the BAO model (black), the no-wiggles model (blue), and zero-baryon model (purple). The no-wiggles model has just the BAO peak smoothed out, whereas the zero-baryon model has a different global shape. We also plot the lognormal simulations mean and error bars (red) which shows simulations are very precise. Finally we plot two other BAO models by changing $\Omega_m h^2=0.1423$ (orange) and $\Omega_m h^2=0.1223$ (green). Increasing $\Omega_m h^2$ reduces the baryonic effects such as the BAO peak, whereas decreasing $\Omega_m h^2$ amplifies these effects.}
\label{cf_plot} 
\end{figure}

The lognormal simulations correspond to parameters $\Omega_m h^2=0.1323$, $\alpha=1$, and $b=2.5$. In figure \ref{cf_plot} we plot the correlation function of the simulations, for the two corresponding no-BAO models and for different values of $\Omega_m h^2$. The no-wiggles model has only the BAO peak smoothed out, whereas the zero-baryon model (with $\Omega_b=0$) has a different global shape. Among BAO models, $\Omega_m h^2$ controls the proportion of baryons in the total matter. When $\Omega_m h^2$ increases, the proportion of baryons decreases, and baryonic effects such as the BAO peak are reduced. On the contrary when $\Omega_m h^2$ decreases, baryonic effects are amplified.

\subsection{Verification of the Gaussianity of $\hat{\xi}$}
\label{verif_gaussian_simu}
In this section we verify the Gaussian hypothesis $\hat{\xi} \thicksim \mathcal{N} \left(\xi_{BAO},C\right)$ on our lognormal simulations. Even if we do not expect large differences, the next step would be to verify it on $N$-body simulations, which are more realistic.

First we look at the $\chi^2_{BAO}$ statistic 
\begin{small}
\begin{eqnarray*}
\chi_{BAO}^2 	& = &  \left \langle \hat{\xi}-\xi_{BAO}, C^{-1} (\hat{\xi}-\xi_{BAO}) \right \rangle \\
 			& = & \sum_{1\leq i,j\leq n}  \left[ \hat{\xi}(r_i)-\xi_{BAO}(r_i) \right] C^{-1}_{i,j}  \left[ \hat{\xi}(r_j)-\xi_{BAO}(r_j) \right]
\end{eqnarray*}
\end{small}

With the Gaussian hypothesis $\hat{\xi} \thicksim \mathcal{N}\left(\xi_{BAO},C\right)$, $\chi_{BAO}^2$ follows a chi-square distribution with $n$ degrees of freedom
\begin{equation}
\chi_{BAO}^2 \thicksim \chi^2_n
\end{equation}

\begin{figure}  
\plotone{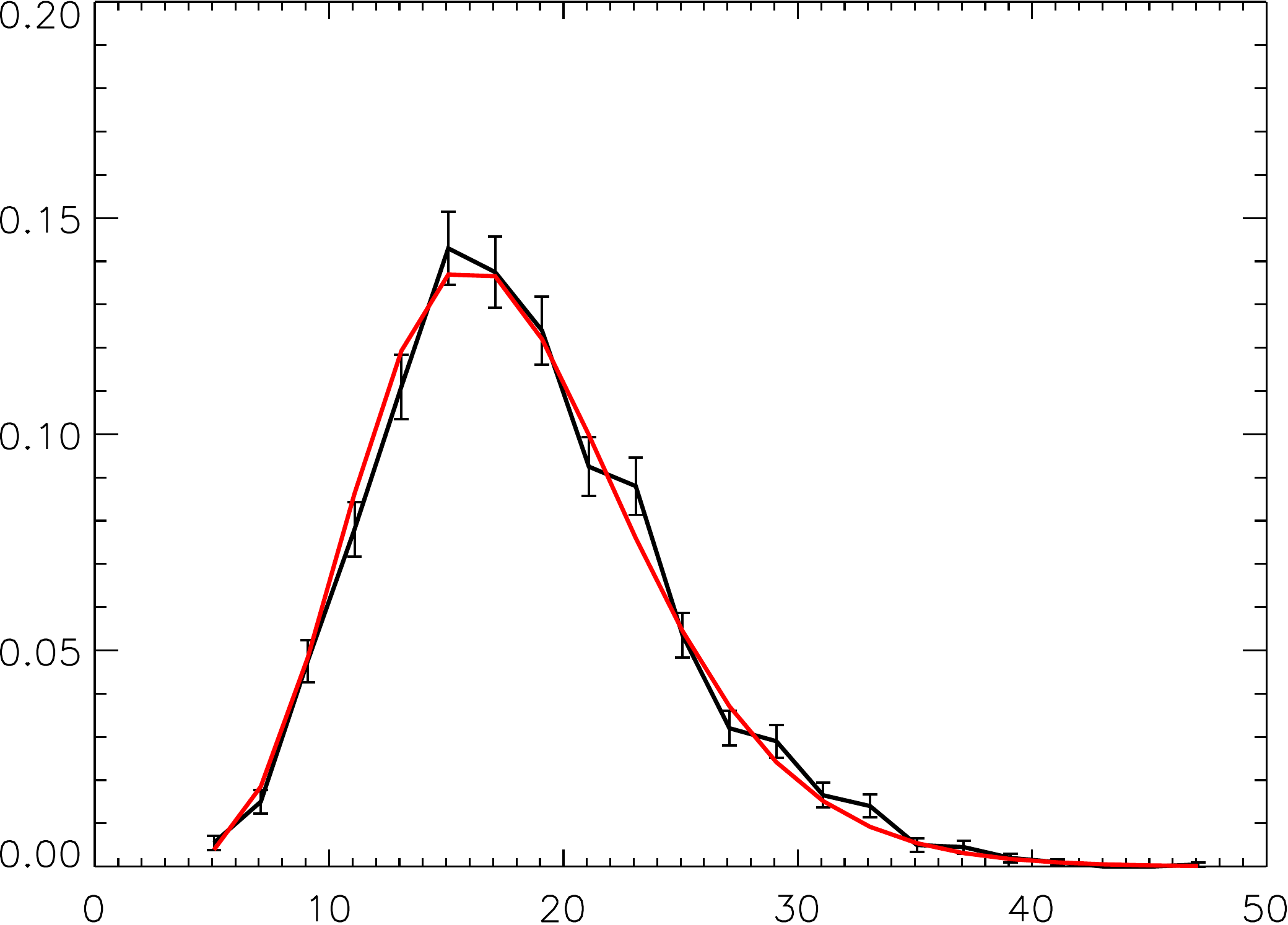} 
      \caption{Estimated pdf of $\chi^2_{BAO}$ (black) using the histogram on 2000 lognormal simulations and pdf of a $\chi^2_{18}$ distribution (red). Error bars give the Poisson error in the estimate due to finite number of simulations.}
\label{chi2_histo} 
\end{figure}

We compare the histogram of $\chi^2_{BAO}$ on our lognormal simulations to the probability density function (pdf) of a $\chi^2_n$ variable where $n=18$.  We show figure \ref{chi2_histo} the very good agreement between the two distributions. \\

For the next tests, we look at wavelet methods of section \ref{wavelet_section}. In these methods we obtain a wavelet response $S_w(R,s)$ and a $Z$-score $Z_w(R,s)$ for every parameter $(R,s)$
\begin{eqnarray*}
S_w(R,s)  & = & \left \langle w(R,s), \hat{\xi} \right \rangle = \sum^n_{i=1} w_i(R,s) \hat{\xi}(r_i) \\
Z_w(R,s) & = &\frac{S_w(R,s)}{\sigma_w(R,s)} 
\end{eqnarray*}
We only consider the mexican hat filter with parameters $R=113.6 \, h^{-1} \mbox{Mpc}$, $s=20 \, h^{-1} \mbox{Mpc}$ and the BAOlet filter with parameters $R=116\, h^{-1} \mbox{Mpc}$, $s=36 \, h^{-1} \mbox{Mpc}$. These parameters maximize the $Z$-score obtained on the data measurement, in the respective studies \cite{Tia11} and \cite{Arn11}. In order to obtain $Z_w$ on our simulations, we compute the noise $\sigma_w$ of $S_w$ using the covariance matrix of the simulations
\begin{equation}
\sigma_w=\sqrt{ \left \langle w, C w \right \rangle}
\end{equation}

\begin{figure}  
\plotone{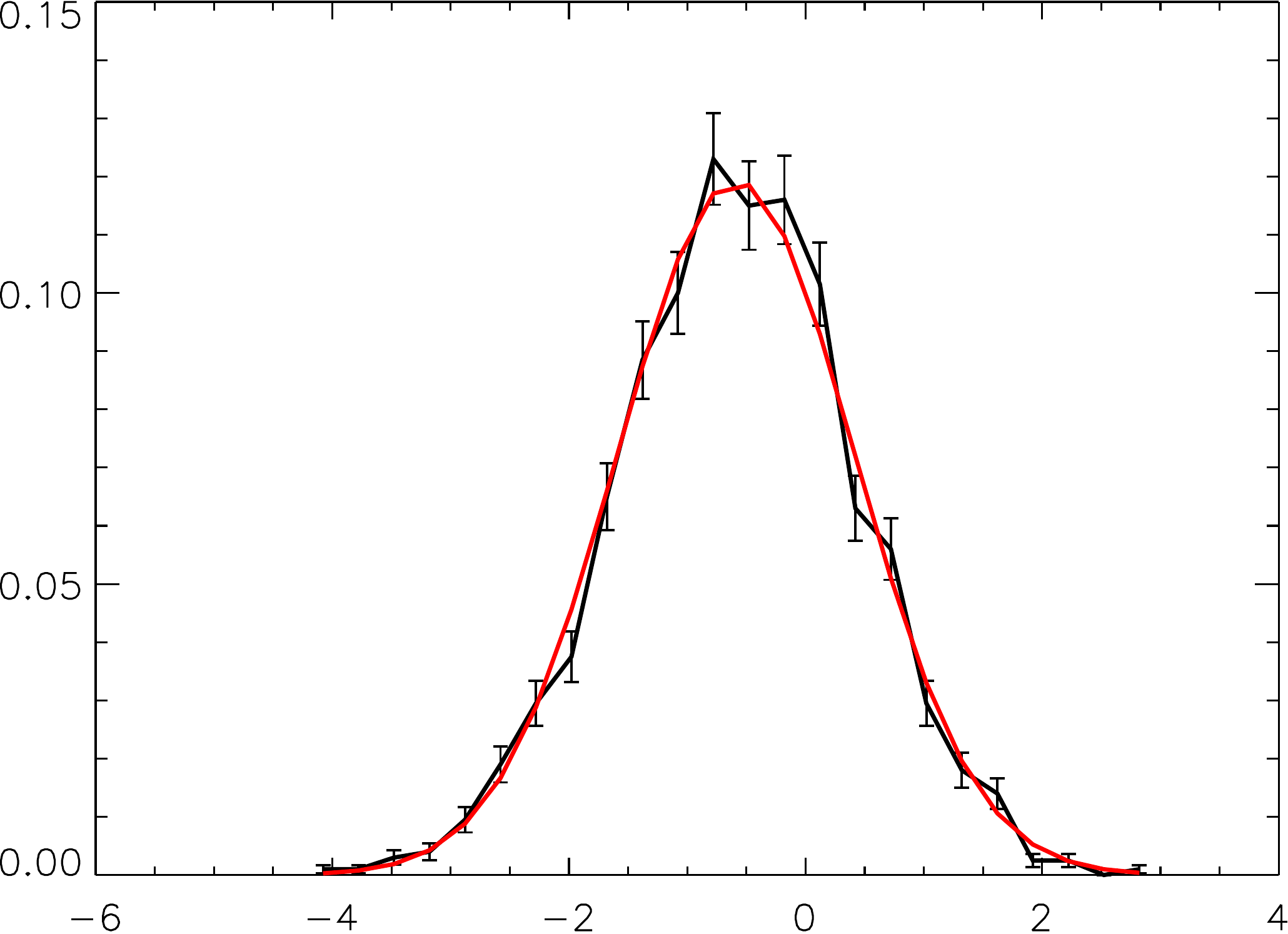} 
      \caption{Estimated pdf of $Z_w$ for the mexican hat filter with parameters $R=113.6 \, h^{-1} \mbox{Mpc}$, $s=20 \, h^{-1} \mbox{Mpc}$ using the histogram on 2000 lognormal simulations (black), and pdf of a standard Gaussian centered on $\overline{Z_w}$ (red). Error bars give the Poisson error in the estimate due to finite number of simulations.}
\label{Zmexhat_histo} 
\end{figure}

\begin{figure}  
\plotone{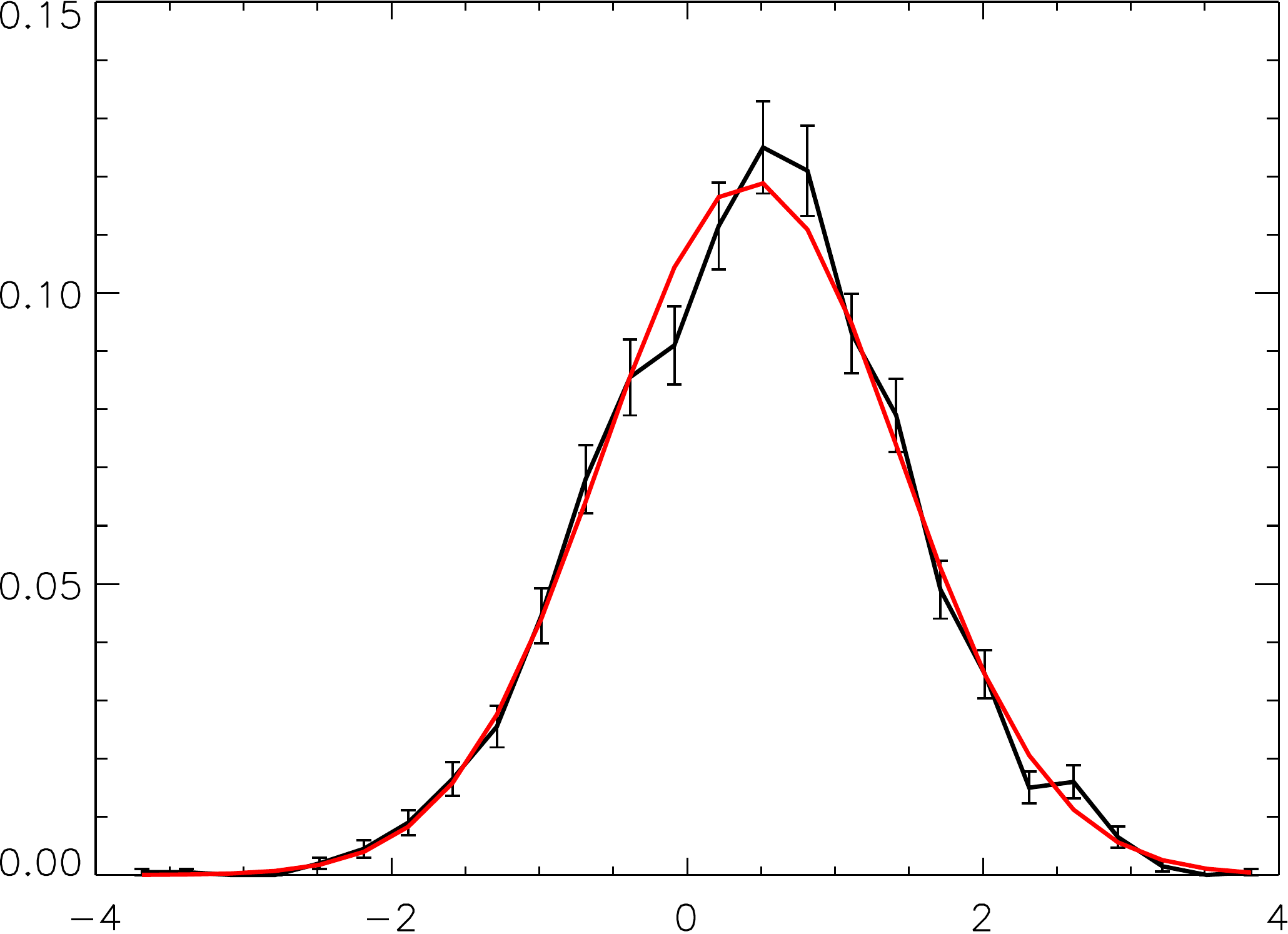} 
      \caption{Estimated pdf of $Z_w$ for the BAOlet filter with parameters $R=116\, h^{-1} \mbox{Mpc}$, $s=36 \, h^{-1} \mbox{Mpc}$ using the histogram on 2000 lognormal simulations (black), and pdf of a standard Gaussian centered on $\overline{Z_w}$ (red). Error bars give the Poisson error in the estimate due to finite number of simulations.}
\label{Zbaolet_histo} 
\end{figure}

With the Gaussian hypothesis $\hat{\xi} \thicksim \mathcal{N}\left(\xi_{BAO},C\right)$, $Z_w$ is Gaussian with mean $\mathbb{E}[Z_w]$ and standard deviation equal to 1. We plot in figures \ref{Zmexhat_histo} and \ref{Zbaolet_histo} the histogram of $Z_w$ on our lognormal simulations, respectively for the mexican hat and for the BAOlet filter. As we already mentioned, there is a small bias between $\overline{Z_w}$ on simulations and $\mathbb{E}[Z_w]$. Here we are only interested in the Gaussianity of $Z_w$ and not in this small bias, so we compare the histogram to the pdf of a Gaussian $\mathcal{N}\left(\overline{Z_w},1 \right)$. Again we find a very good agreement between the simulations and the Gaussian prediction. \\

\subsection{BAO detection with constant covariance matrix}

\subsubsection{Classical $\chi^2$ method}
\label{classical_chi2_simu}
The tested hypotheses are given by 
\begin{eqnarray*}
\mathcal{H}_0 & : &  \exists \, \theta \in \Theta \,\, \mbox{s.t.} \,\, \hat{\xi} \thicksim \mathcal{N} \left( \xi_{noBAO,\theta}, C \right) \\
\mathcal{H}_1 & : &  \exists \, \theta \in \Theta \,\, \mbox{s.t.} \,\, \hat{\xi} \thicksim \mathcal{N} \left( \xi_{BAO,\theta}, C \right) 
\end{eqnarray*}
The Gaussian hypothesis on $\hat{\xi}$ is well justified (at least for our lognormal simulations) as we have seen in section \ref{verif_gaussian_simu}. The chi-square quantities are functions of $\theta$
\begin{eqnarray*}
\chi_{BAO,\theta}^2 &= & \left \langle \hat{\xi}-\xi_{BAO,\theta}, C^{-1} (\hat{\xi}-\xi_{BAO,\theta}) \right \rangle\\
\chi_{noBAO,\theta}^2 &= &  \left \langle \hat{\xi}-\xi_{noBAO,\theta}, C^{-1} (\hat{\xi}-\xi_{noBAO,\theta}) \right \rangle
\end{eqnarray*}

An extended model of correlation function is implicitly defined, which mixes the BAO and the no-BAO models, e.g.
\begin{equation*}
\xi_{\beta,\theta} = \beta \, \xi_{BAO,\theta} + (1-\beta) \, \xi_{noBAO,\theta}  \\
\end{equation*}

Let us write as before
\begin{eqnarray*}
\Delta \chi^2 & = & \min_\theta \chi_{noBAO,\theta}^2-\min_\theta \chi_{BAO,\theta}^2 \\
\Delta \chi^2_{global} & = &\min_{\theta} \chi_{noBAO,\theta}^2-\min_{\beta,\theta} \chi_{\beta,\theta}^2 \\
\end{eqnarray*}

The basic assumption of the method is that $\Delta \chi^2_{global}$ follows a $\chi^2_1$ distribution under $\mathcal{H}_0$. Since we have $\Delta \chi^2  \le \Delta \chi^2_{global}$ by construction, a conservative estimate of the significance is given by $\sqrt{\Delta \chi^2}.\sigma$ when $\Delta \chi^2 \ge 0$.

We recall that $\Delta \chi^2_{global} \thicksim \chi^2_1$ is subject to the assumption that the spaces of model correlation functions $(\xi_{noBAO,\theta})_{\theta \in \Theta}$ and $(\xi_{\beta,\theta})_{\beta \in \mathbb{R},\theta \in \Theta}$ are affine. Since it is not easy to verify we want to test with our simulations that we have indeed $\Delta \chi^2_{global} \thicksim \chi^2_1$ under $\mathcal{H}_0$.

For a model $\theta$ in $\mathcal{H}_0$, we generate realizations as
\begin{equation}
C^{1/2} g+\xi_{noBAO,\theta}
\label{generate}
\end{equation}

where $g$ is a standard multivariate Gaussian. For each realization, we find the best-fit model by testing all the $\Omega_m h^2$ and $\alpha$ values on a grid. The remaining parameters, which are the bias $b^2$ and the parameter $\beta$, are found analytically for the best-fit. 

For the grid $(\Omega_m h^2,\alpha)$ we take $\Omega_m h^2 \in [0.0423,0.2923]$ with grid step 0.005 and $\alpha \in [0.5,1.5]$ with grid step 0.01. We also allow any $b^2 \ge 0$ and $\beta$. We test two models in $\mathcal{H}_0$ with values $\Omega_m h^2=0.1323$, $\alpha=1.0$ as in the lognormal simulations, and with other values $\Omega_m h^2=0.0823$, $\alpha=0.9$. Each time we generate 10 000 realizations using equation (\ref{generate}) to estimate the distribution of $\Delta \chi^2_{global}$ and $\Delta \chi^2$. We show in tables \ref{chi2_assumption1} and \ref{chi2_assumption2} for different thresholds $t$, the $p$-value and corresponding significance for a $\chi^2_1$ variable, for $\Delta \chi^2_{global}$, and for $\Delta \chi^2$. Our results show that the assumption $\Delta \chi^2_{global} \thicksim \chi^2_1$ is clearly wrong. In particular, the mean of $\Delta \chi^2_{global}$ for the two different models are respectively 2.85 and 2.23, whereas the mean of a $\chi^2_1$ variable is equal to 1.

In both cases one grossly overestimates the significance when identifying $\Delta \chi^2_{global}$ with a $\chi^2_1$ distribution. The fact that the classical $\chi^2$ method is conservative because it uses the value of $\Delta \chi^2$ instead of $\Delta \chi^2_{global}$ can compensate this overestimation. For one $\mathcal{H}_0$ model, identifying $\Delta \chi^2$ with a $\chi^2_1$ distribution still gives a small overestimation of the significance. For the other $\mathcal{H}_0$ model, it gives an underestimation of the significance.

Let us stress that these significances are not strictly speaking the significances of the BAO detection, but only the rejection of particular $\mathcal{H}_0$ models. Indeed the BAO detection consists in the rejection of all $\mathcal{H}_0$ models simultaneously.

\begin{table}[htbp]
\scriptsize
\caption{\label{chi2_assumption1}}
\begin{center}
\begin{tabular}{llll} 
\tableskip\tableline\tableline\tableskip
			     &  \multicolumn{1}{c}{$\chi^2_1$}& \multicolumn{1}{c}{$\Delta \chi^2_{global}$} & \multicolumn{1}{c}{$\Delta \chi^2$} \\ 
\tableskip\tableline\tableline\tableskip
$P(X \!\ge\! 1.0)$    	& $0.32\, (1\sigma$)					 & $0.81\, (0.25\sigma$) 				& $0.39\, (0.85\sigma$) \\
$P(X  \!\ge\! 2.25)$   	& $0.13\, (1.5\sigma$) 				& $0.51\, (0.65\sigma$)				 & $0.18\, (1.35\sigma$)\\
$P(X  \!\ge\! 4.0)$     	& $4.5 \! \times \!10^{-2}\, (2\sigma$) 	& $0.23\, (1.2\sigma$) 				& $6.8 \! \times \! 10^{-2}\, (1.8\sigma$)\\
$P(X  \!\ge\! 6.25)$  	 & $1.2 \! \times \!10^{-2}\,  (2.5\sigma$) 	& $8.3 \! \times \! 10^{-2}\, (1.75\sigma$) 	& $2.1\! \times \! 10^{-2}\, (2.3\sigma$)\\
$P(X  \!\ge\! 9.0)$	& $2.7 \! \times \! 10^{-3}\, (3\sigma$) 	& $1.9 \! \times \! 10^{-2}\, (2.35\sigma$)	& $4.3 \! \times \! 10^{-3}\,  (2.85\sigma$)\\ 
\tableskip\tableline\tableline\tableskip
\end{tabular}
\end{center}
NOTES.---%
$p$-values and corresponding significances for different distributions and for the rejection of the particular $\mathcal{H}_0$ model with $\Omega_m h^2=0.1323$, $\alpha=1.0$. We show the $\chi^2_1$ distribution and the $\Delta \chi^2_{global}$, $\Delta \chi^2$ distributions. The assumption $\Delta \chi^2_{global} \thicksim \chi^2_1$ is wrong with a significance that is grossly overestimated. The fact that the classical $\chi^2$ method uses $\Delta \chi^2$ instead of $\Delta \chi^2_{global}$ compensates the overestimation. For the rejection of this $\mathcal{H}_0$ model, there is still a small overestimation of the significance if we identify $\Delta \chi^2$ with a $\chi^2_1$ distribution.
\end{table}

\begin{table}[htbp]
\scriptsize
\caption{\label{chi2_assumption2}}
\begin{center}
\begin{tabular}{llll} 
\tableskip\tableline\tableline\tableskip
			     &  \multicolumn{1}{c}{$\chi^2_1$}& \multicolumn{1}{c}{$\Delta \chi^2_{global}$} & \multicolumn{1}{c}{$\Delta \chi^2$} \\ 
\tableskip\tableline\tableline\tableskip
$P(X \! \ge \! 1.0)$     	&  $0.32\, (1\sigma$)			  		 & $0.66\, (0.45\sigma$)				 & $8.0 \! \times \!10^{-2}\, (1.75\sigma$) \\
$P(X \! \ge \!  2.25)$  	&  $0.13\, (1.5\sigma$) 				 & $0.37\, (0.9\sigma$)				 & $4.2 \! \times \!10^{-2}\, (2.05\sigma$)\\
$P(X \! \ge \!  4.0)$    	& $4.5 \! \times \!10^{-2}\, (2\sigma$)		 & $0.17\, (1.4\sigma$) 				 & $1.6 \! \times \! 10^{-2}\, (2.4\sigma$)\\
$P(X \! \ge \!  6.25)$  	& $1.2 \! \times \!10^{-2}\,  (2.5\sigma$)	& $5.5 \! \times \! 10^{-2}\, (1.9\sigma$)  	& $5.7\! \times \! 10^{-3}\, (2.75\sigma$)\\
$P(X \! \ge \!  9.0)$   	& $2.7 \! \times \! 10^{-3} \,(3\sigma$)    	& $1.3 \! \times \! 10^{-2}\, (2.5\sigma$)	& $1.8 \! \times \! 10^{-3}\,  (3.1\sigma$)\\ 
\tableskip\tableline\tableline\tableskip
\end{tabular}
\end{center}
NOTES.---%
Same as table \ref{chi2_assumption1} for the rejection of the particular $\mathcal{H}_0$ model with $\Omega_m h^2=0.0823$, $\alpha=0.9$. Again the assumption $\Delta \chi^2_{global} \thicksim \chi^2_1$ is wrong with a significance that is grossly overestimated. The fact that the classical $\chi^2$ method uses $\Delta \chi^2$ instead of $\Delta \chi^2_{global}$ compensates the overestimation. For the rejection of this $\mathcal{H}_0$ model, the significance becomes underestimated if we identify $\Delta \chi^2$ with a $\chi^2_1$ distribution.\\ 
\end{table}

When using zero-baryon models for $\mathcal{H}_0$ we find the same qualitative results. We find that $\Delta \chi^2_{global}$ is very different from a $\chi^2_1$ variable, and that identifying $\Delta \chi^2$ with a $\chi^2_1$ distribution is also wrong, leading to either an overestimation or an underestimation of the significance for the rejection of particular $\mathcal{H}_0$ models.

The assumption $\Delta \chi^2_{global} \thicksim \chi^2_1$ is broken because the spaces of model correlation functions $(\xi_{noBAO,\theta})_{\theta \in \Theta}$ and $(\xi_{\beta,\theta})_{\beta \in \mathbb{R},\theta \in \Theta}$ are not affine. It is easy to see for example that no-BAO correlations are more degenerate than BAO correlations with respect to the three parameters. So for a given range of parameters, the space of BAO correlations is larger than the space of no-BAO correlations, which tends to increase the values of $\Delta \chi^2_{global}$ and $\Delta \chi^2$. In this case, spaces of correlations fail to be affine because of their limited extent, which itself is due to the limited range of parameters.

As we saw in section \ref{modified_chi2_section}, one needs to consider the "worst-case" $\mathcal{H}_0$ model to obtain the correct significance. For a realization value $\Delta \chi^2=x$ the $p$-value is given by
\begin{equation}
\label{worstcase_pvalue}
p(x)=\max_{\theta \in \Theta} P(\Delta \chi^2 \ge x \, | \, \mathcal{H}_0, \theta)
\end{equation} 

We saw in tables \ref{chi2_assumption1} and \ref{chi2_assumption2} that the significance can be either overestimated or underestimated when rejecting particular $\mathcal{H}_0$ models, if we identify $\Delta \chi^2$ with a $\chi^2_1$ variable. If it is overestimated for a $\mathcal{H}_0$ model with parameter $\theta$ and for $\Delta \chi^2=x \ge 0$, it means
\begin{equation}
P(\Delta \chi^2 \ge x \, | \, \mathcal{H}_0, \theta) > p_{\chi^2_1}(x) =2\Phi(-\sqrt{x})
\end{equation} 

The consequence is that the classical $\chi^2$ method overestimates the significance of the BAO detection. Indeed when using equation (\ref{worstcase_pvalue}) for determining the significance of the full $\mathcal{H}_0$ rejection, we get
\begin{equation}
p(x) > p_{\chi^2_1}(x) =2\Phi(-\sqrt{x})
\end{equation} 

When considering varying covariance matrices, the estimate of the significance by the classical $\chi^2$ method could even be more wrong (see section \ref{chi2_limitations} and section \ref{detect_varcov_simu}).

\subsubsection{$\Delta l$ method}
\label{modified_chi2_simu}
In this section we test the modified version of the $\chi^2$ method that we proposed in section \ref{modified_chi2_section}, which we called the $\Delta l$ method. We still consider a constant covariance matrix, i.e. the hypotheses
\begin{eqnarray*}
\mathcal{H}_0 & : &  \exists \, \theta \in \Theta \,\, \mbox{s.t.} \,\, \hat{\xi} \thicksim \mathcal{N} \left( \xi_{noBAO,\theta}, C \right) \\
\mathcal{H}_1 & : &  \exists \, \theta \in \Theta \,\, \mbox{s.t.} \,\, \hat{\xi} \thicksim \mathcal{N} \left( \xi_{BAO,\theta}, C \right) 
\end{eqnarray*}
There are two modifications that we proposed compared to the classical $\chi^2$ method. One of the modifications consists in replacing $\Delta \chi^2$ by $\Delta l$
\begin{equation}
\Delta l  = -2\left[\max_\theta \ln \left( \mathcal{L}_{noBAO,\theta} \right) - \max_\theta \ln \left( \mathcal{L}_{BAO,\theta}\right) \right]\\
\end{equation}

We are still in the case of a constant covariance, so the two statistics $\Delta \chi^2$ and $\Delta l$ are equal. We will only see the effect of this change in section \ref{detect_varcov_simu} when we consider varying covariance matrices. 

We also modify the procedure for computing the significance, so that we obtain a correct value as in equation (\ref{pvalue_modified_chi2}) or (\ref{worstcase_pvalue}). For a realization value $\Delta l=x$ the $p$-value is given by
\begin{equation}
p(x)=\max_{\theta \in \Theta} P(\Delta l \ge x \, | \, \mathcal{H}_0, \theta)
\end{equation} 

Let us define a cumulative distribution function corresponding to this $p$-value
\begin{equation}
F_0(x)=p(x)=\max_{\theta \in \Theta} P(\Delta l \ge x \, | \, \mathcal{H}_0, \theta)
\label{f0}
\end{equation} 

So the method requires to precompute $F_0$ in order to obtain the $p$-value $p(x)$ for a given measurement $\Delta l=x$. Here we consider the range of parameters  $\Omega_m h^2 \in [0.1023,0.1623]$ with grid step 0.015, $\alpha \in [0.8,1.2]$ with grid step 0.05, and $b^2 \in [4,9]$ with grid step 0.25. The grid is not very fine because the computation time is proportional to the square of the grid size, so it increases very rapidly. We tested to refine the grid for each parameter and found agreements at a few percents, which is enough for our purpose here. 

For each model on the grid we generate 10.000 realizations with equation (\ref{generate}) as before, and compute the statistic $\Delta l$. The fact that there are only 10.000 realizations for each $\mathcal{H}_0$ model does not enable to quantify $p$-values smaller than approximately $10^{-4}$, i.e. significances higher than $3.85\sigma$. Note that unlike in the $\chi^2$ method, the imprecision here is only computational and limited to high significance. In particular when working with a data measurement $\hat{\xi}$, the imprecision only occurs when the significance is high and the BAO detection is already clear.

We perform this procedure both for the $\mathcal{H}_0$ hypothesis and for the $\mathcal{H}_1$ hypothesis. With $\mathcal{H}_0$ realizations we compute the function $F_0$, and we get the significance obtained with $\mathcal{H}_1$ realizations using equation (\ref{f0}). Since $\mathcal{H}_1$ is composite, the distribution of $\Delta l$ is not well-defined under $\mathcal{H}_1$. For example we cannot speak about the expected significance obtained under $\mathcal{H}_1$. Here we simply consider the expected significance for every $\mathcal{H}_1$ model, that we average over all models. This is actually equivalent to the expected significance obtained under $\mathcal{H}_1$ when adding a constant prior $p(\theta)$ in the hypothesis.

We obtain an average significance of $2.11\sigma$ with this procedure. On the other hand, when estimating the significance by $\sqrt{\Delta \chi^2}.\sigma$ as in the classical $\chi^2$ method, we obtain an average of $2.33\sigma$ (we take the convention that $\Delta \chi^2 \le 0$ corresponds to $0\sigma$). Let us see the effect of the imprecision at high significance by only considering realizations under the limit of $3.85\sigma$. In this case we obtain an average significance of $1.92\sigma$ with the modified procedure, and an average of $\sqrt{\Delta \chi^2}$ equal to $2.0$. Note that this does not mean that the $\chi^2$ method is better. Indeed it uses the same statistic, so this only means that the significance is overestimated.

Finally we repeat the same computations using zero-baryon models for $\mathcal{H}_0$. In this case we expect a larger significance because zero-baryon models not only lack the BAO feature, but also have a different global shape than baryonic models. The expected significance is higher, so we use a higher number of realizations equal to 50.000 for every $\mathcal{H}_0$ model.  When using the rigorous procedure for estimating the significance, we obtain an average of 2.34$\sigma$. On the other hand, when using $\sqrt{\Delta \chi^2}$ we obtain an average of $2.9$. Here the large difference is mostly due to the imprecision at high significance. Indeed if we restrict to significances under the limit of $4.25\sigma$ corresponding to this number of realizations, we obtain an average significance of $2\sigma$ with our modified procedure, and an average of $\sqrt{\Delta \chi^2}$ equal to 2.21. 

\subsection{BAO detection with varying covariance matrix}
\label{detect_varcov_simu}
We finally consider the general case where the covariance matrix depends on cosmological parameters
\begin{eqnarray*}
\mathcal{H}_0 & : &  \exists \, \theta \in \Theta \,\, \mbox{s.t.} \,\, \hat{\xi} \thicksim \mathcal{N} \left( \xi_{noBAO,\theta}, C_{noBAO,\theta} \right) \\
\mathcal{H}_1 & : &  \exists \, \theta \in \Theta \,\, \mbox{s.t.} \,\, \hat{\xi} \thicksim \mathcal{N} \left( \xi_{BAO,\theta}, C_{BAO,\theta} \right) 
\end{eqnarray*}

Our goal is only to illustrate the effect of a varying covariance instead of a constant covariance. So we consider a simple example where the covariance matrix only depends on the amplitude parameter $b$
\begin{equation}
C_{noBAO,\theta}=C_{BAO,\theta}=\left(\frac{b}{b_0}\right)^4 C
\end{equation}
with $b_0=2.5$ the simulation value. 

We apply the $\Delta l$ method with the same procedure as in section \ref{modified_chi2_simu}, except we take into account variations of the covariance matrix in the likelihoods and when generating $\mathcal{H}_0$, $\mathcal{H}_1$ realizations using equation (\ref{generate}). In this case, the $\Delta l$ statistic is different from the $\Delta \chi^2$ statistic, and is computed using equations (\ref{chi2_extended1}), (\ref{chi2_extended2}), (\ref{delta_l}), (\ref{logL_extended1}), and (\ref{logL_extended2}).

This time we obtain an average significance of $1.96\sigma$ under $\mathcal{H}_1$, which is a bit lower than for a constant covariance. When using the $\Delta \chi^2$ statistic we obtain an average of $1.59\sigma$ under $\mathcal{H}_1$ with our modified procedure. This justifies our choice of replacing $\Delta \chi^2$ by $\Delta l$, which can be thought as a generalized likelihood ratio (see section \ref{modified_chi2_section}). Using the classical $\chi^2$ method, we obtain an average of $\sqrt{\Delta \chi^2}$ equal to 2.32. In this case the estimate given by the classical $\chi^2$ method is very far from the correct significance of $1.59\sigma$. As we already mentioned in section \ref{chi2_limitations}, the classical $\chi^2$ method cannot be used in the case of a varying covariance matrix. 

We verify that these conclusions are not due to the imprecision at high significance of our procedure. When considering only realizations under the limit of $3.85\sigma$, our modified procedure gives average significances of $1.89\sigma$ for $\Delta l$ and $1.52\sigma$ for $\Delta \chi^2$, and the average of $\sqrt{\Delta \chi^2}$ is 2.20.

These results show how the BAO detection is dependent on the tested hypotheses $\mathcal{H}_0$, $\mathcal{H}_1$, and the choice of the statistic. \\ \\

\subsection{Effect of varying covariance matrix on cosmological parameters constraints}
\label{constraints_varcov_simu}
Finally let us see the effect of a varying covariance matrix on parameter constraints. As we saw in section \ref{constraints_section}, we must have a prior $p(\theta)$ for the posterior $p(\theta \, | \, \hat{\xi})$ to be well-defined. We consider a constant $p(\theta)$ so that the constraints only come from the measurement $\hat{\xi}$. Then the posterior $p(\theta \, | \, \hat{\xi})$ is given by the likelihood 
\begin{equation}
p(\theta \, | \, \hat{\xi}) \propto \mathcal{L}_{BAO,\theta}
\end{equation}

Changing the covariance matrix is equivalent to changing the likelihood function. Let us see the effect for a given measurement $\hat{\xi}$. We use for the illustration the expected correlation function of the lognormal simulations $\hat{\xi}=\xi_{BAO,\theta}$ with $\theta=(\Omega_mh^2,\alpha,b)=(0.1323,1.0,2.5)$. 

\begin{figure}[htbp]
\plotone{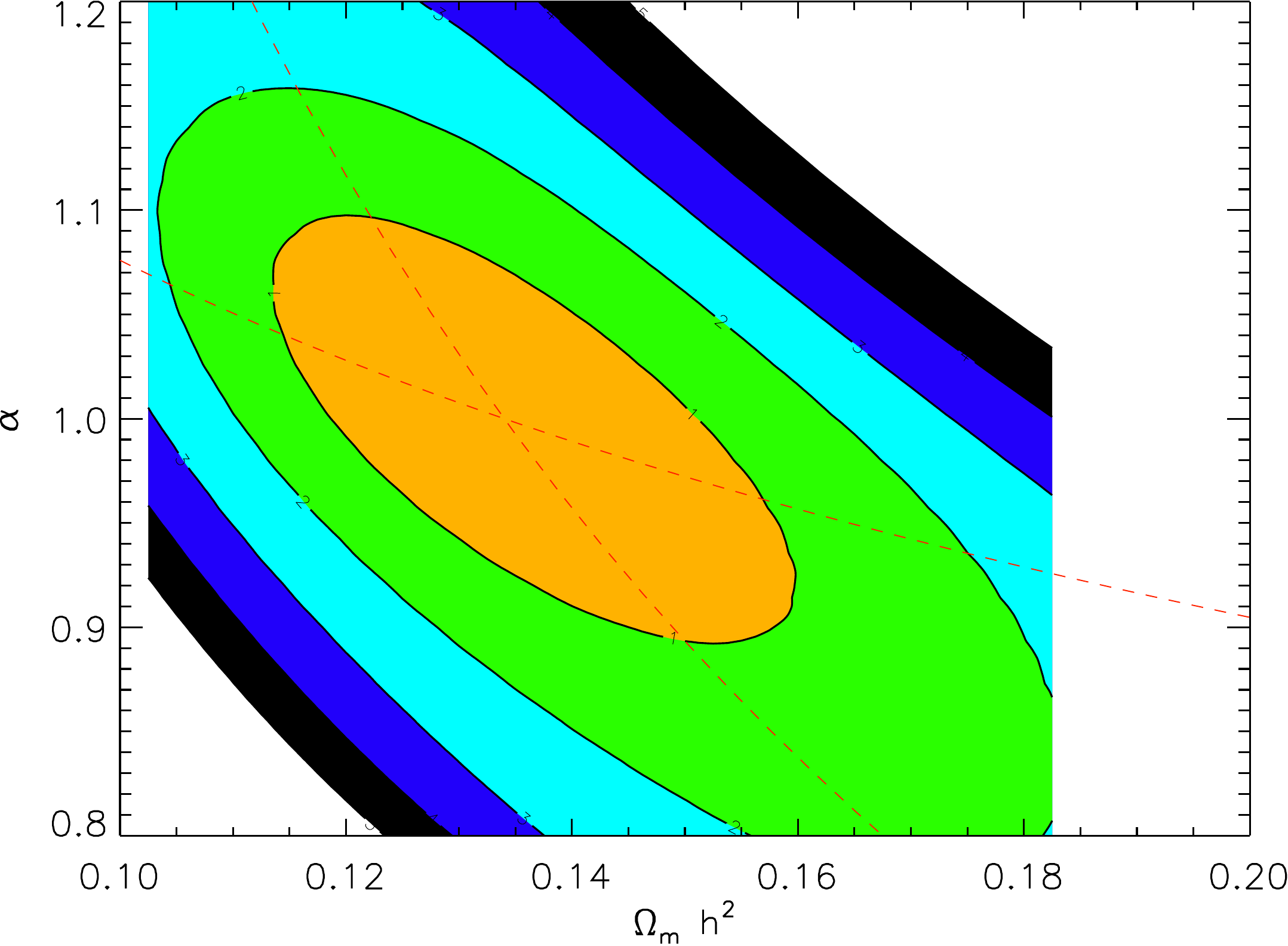} 
      \caption{Posterior $p(\Omega_m h^2,\alpha \, | \, \hat{\xi})$ in the case of constant covariance matrix, with $\hat{\xi}=\xi_{BAO,\theta}$ and $\theta=(\Omega_m h^2,\alpha,b)=(0.1323,1.0,2.5)$ for the illustration. We plot the $1\sigma$ to $5\sigma$ confidence regions with the approximation that $p$ is a 2-dimensional Gaussian. They correspond respectively to $-2\ln(p)=-2\ln(p_{max})+2.29,6.16,11.81,19.32,28.74$ (see section "Confidence Limits on Estimated Model Parameters" in \cite{Pre07}). We see deviations to a Gaussian posterior because the contours are not totally elliptical and symmetrical. We also plot the lines of constant apparent horizon at matter-radiation equality $\alpha\, \Omega_mh^2$, and constant apparent sound horizon $\alpha \, (\Omega_mh^2)^{0.25}$. As expected the degeneracy direction of the constraints lies in between the two lines. }
\label{constraints1} 
\end{figure}

\begin{figure}[htbp]
\plotone{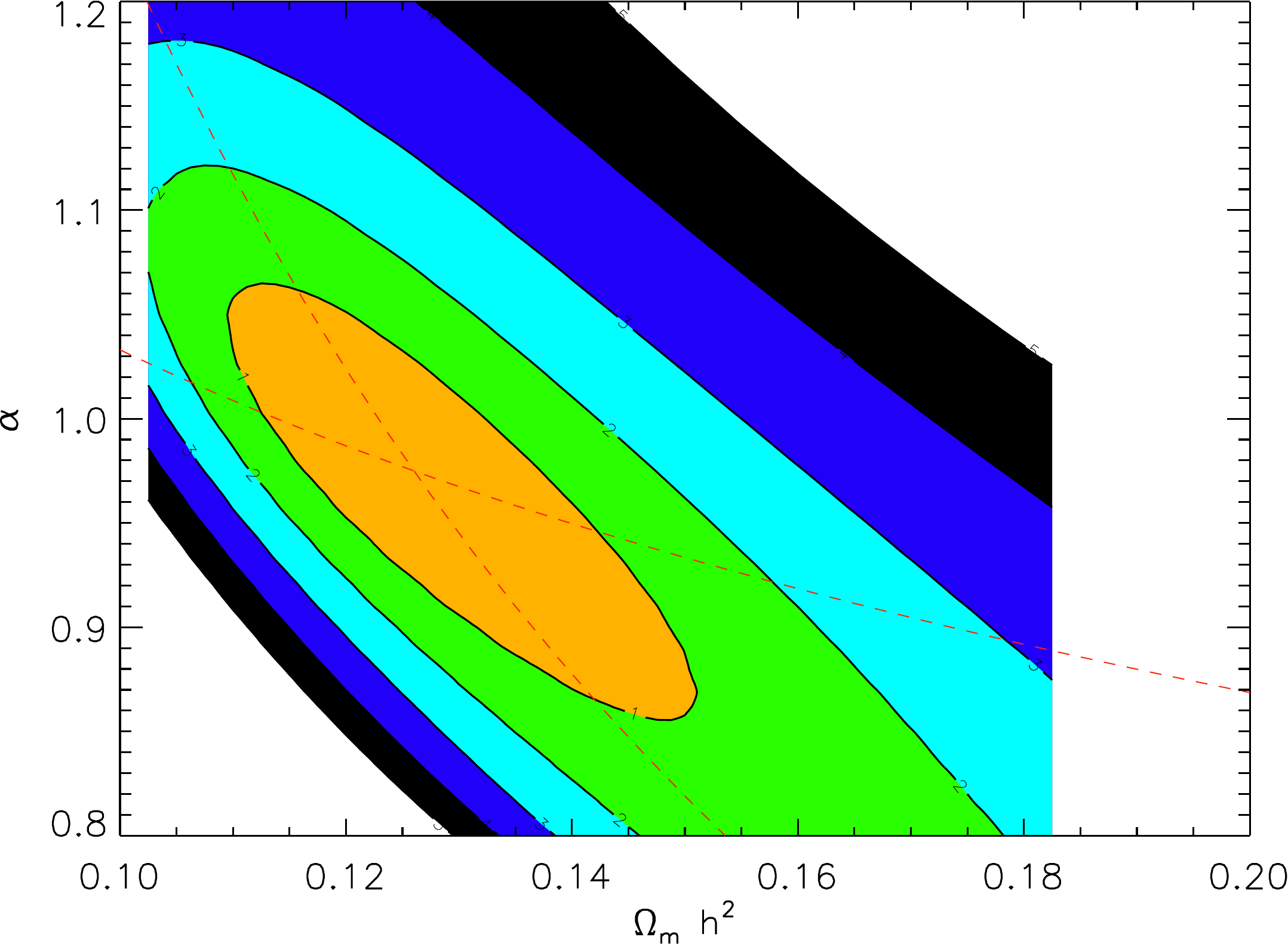} 
      \caption{Same as figure \ref{constraints1} but for varying covariance matrices $C_{BAO,\theta}=(\frac{b}{b_0})^4C$. In this case the posterior $p(\Omega_m h^2,\alpha \, | \, \hat{\xi})$ is much farther from a 2-dimensional Gaussian. The degeneracy direction is as poorly constrained as in figure \ref{constraints1}. However the orthogonal direction is better constrained. Overall the 2-dimensional constraints on $(\Omega_mh^2,\alpha)$ are better for this particular realization. \\ \\}
\label{constraints2} 
\end{figure}

We compute the posterior $p(\Omega_mh^2,\alpha \, | \, \hat{\xi})$ after marginalizing over the amplitude $b^2$ with $b^2 \in [4,9]$. We plot the results in figure \ref{constraints1} and \ref{constraints2} respectively for a constant covariance matrix and for a varying covariance matrix. We also plot two lines of constant apparent horizon at matter-radiation equality $\alpha\, \Omega_mh^2$, and constant apparent sound horizon $\alpha\, (\Omega_mh^2)^{0.25}$. These would be degeneracy lines if we focused respectively on small scales and on the BAO scale. As expected the degeneracy direction for the constraints lies in between the two lines.

First we notice that the posterior $p(\Omega_mh^2,\alpha \, | \, \hat{\xi})$ is much farther to a 2-dimensional Gaussian for a varying covariance matrix than for a fixed covariance matrix.

Constraints on each parameter $\Omega_mh^2$ and $\alpha$ are obtained after marginalizing on the other parameter. We obtain different constraints in the two cases, with a small shift in the maxima of the posteriors. For a constant covariance we obtain $\Omega_mh^2=0.134 \pm 0.015$ and $\alpha= 0.995\pm 0.070$. For a varying covariance we obtain $\Omega_mh^2= 0.126 \pm 0.014$ and $\alpha= 0.976 \pm 0.070$.

Constraints on individual parameters are dominated by the degeneracy direction of the 2-dimensional posterior. In both cases this direction is approximately as poorly constrained. However the orthogonal direction is better constrained in the case of varying covariance. Overall the 2-dimensional constraints on $(\Omega_mh^2,\alpha)$ are better for this particular example of correlation function.

\section{Conclusions}
\label{conclusion}
We have presented different methods for BAO detection, and for each of them we detailed the tested $\mathcal{H}_0$, $\mathcal{H}_1$ hypotheses and the underlying assumptions. We show in table \ref{summary} a summary with pros and cons for each method.

A first type of methods is based on wavelet filtering. Their main advantage is that they are mildly model-dependent, and mainly sensitive to the BAO feature in the correlation function. Thus they are only weakly affected by modeling errors. The price to pay is that they are outperformed by some model-dependent methods when $\mathcal{H}_0$ and $\mathcal{H}_1$ are well modeled.

Other methods are fully model-dependent. They assume that the measurement $\hat{\xi}$ is Gaussian, which is well verified on our simulations. They also forbid a too small binning since they can be unstable and use the inverse of the covariance matrix $C$ of $\hat{\xi}$. However this does not cause too much loss of information.
 
Among these methods, the most often used is the classical $\chi^2$, based on the $\chi^2$ statistic. We found that it only gives a rough estimate for the significance of the BAO detection, and more precisely an overestimation. This comes from the method assumption that spaces of model correlation functions are affine, which is not verified in practice. As a consequence, the significance of the rejection of some particular  $\mathcal{H}_0$ models is overestimated. Since the rejection of the full $\mathcal{H}_0$ hypothesis (the BAO detection) is based on the "worst-case" $\mathcal{H}_0$ model, its significance is also overestimated. Moreover the estimate of the significance can become more wrong with hypotheses where the covariance matrix is model-dependent.

We proposed to use the $\Delta l$ method, which is a modified version of the classical $\chi^2$ method. We first modify the procedure for obtaining the correct significance. Using simulations, we found that correct significances are indeed lower than estimates of the classical $\chi^2$ method. The price to pay is that the method becomes much more expensive computationally. As a result we cannot use as many $\mathcal{H}_0$ realizations as we want, which causes imprecision at high significance. Yet this limitation is only computational and restricted to the case of high significance where the BAO detection is already clear.

The second modification consists in replacing the $\Delta \chi^2$ statistic by the $\Delta l$ statistic, which coincide for a constant covariance matrix but are different for a varying covariance matrix. We found that the $\Delta l$ statistic gives better results in the case of varying covariance. As we have seen with a simple example, taking into these variations can affect both the BAO detection and cosmological parameter constraints.

In the course of our study we also found that no-wiggles models are rejected at a lower level than zero-baryon models. It agrees with the analysis in \cite{Cab11}, which uses no-wiggles models for $\mathcal{H}_0$ and finds that the BAO peak is rarely detected above $3\sigma$ for current galaxy surveys. This comes from the fact that no-wiggles correlation functions only lack the BAO peak, whereas zero-baryon correlation functions have a global different shape. So there must be a clear distinction between testing the existence of the BAO peak, and testing the existence of baryons.

Let us summarize our main conclusions:
\begin{enumerate}
\item The choice of the hypotheses $\mathcal{H}_0$ and $\mathcal{H}_1$ is important since it affects both the BAO detection and cosmological parameter constraints. To be rigorous one should take into account variations of the covariance matrix. It should also be clear whether one tests the existence of baryons or only the existence of the BAO peak, because the expected results are quite different.
\item We have presented a new method, the $\Delta l$ method, which has two main advantages over the classical $\chi^2$ method. Unlike the latter it provides the correct significance, apart from imprecisions at high significance. It also provides better results in the case of varying covariance matrix.
\end{enumerate}

We plan to apply the $\Delta l$ method for BAO detection in the LRG sample of SDSS DR7. For this we also plan to use more realistic hypotheses $\mathcal{H}_0$ and $\mathcal{H}_1$, by modeling the variations of the covariance matrix $C(\hat{\xi})$. A more realistic $\mathcal{H}_1$ hypothesis would also give more realistic parameter constraints.

\begin{acknowledgements}
      Part of this work was supported by the European Research Council grant ERC-228261. We would like to thank Nicolas Clerc and Vicent Mart\'inez for useful discussions and comments, as well as the anonymous referee for improving the quality of this paper. \\
\end{acknowledgements}

\begin{table*}[htbp]
\footnotesize
\caption{\label{summary}}
\begin{center}
{\sc Hypotheses, pros, and cons of the different BAO detection methods \\}
\begin{tabular}{cllll} 
\tableskip\tableline\tableline\tableskip
 & \multicolumn{2}{l}{Hypotheses}  &   Pros  & Cons \\ 
\tableskip\tableline\tableline\tableskip
 \it{Wavelet}		 	& $\mathcal{H}_0$: & no peak in $\mathbb{E}[\hat{\xi}] $ & \textbf{- Mildly model dependent} $\rightarrow$ & - Outperformed by some model- \\
 \it{methods}     		        & $\mathcal{H}_1$: & peak in $\mathbb{E}[\hat{\xi}] $		 & \;\;\;\,\textbf{robust to modeling errors}	 & \;\;\;\,dependent methods when there \\
                                 		& & & & \;\;\;\,are no modeling errors \\
	                                 	& & & & \\
\it{Classical $\chi^2$} & $\mathcal{H}_0:$ &  $\exists \, \theta \,\, \mbox{s.t.} \,\, \hat{\xi} \thicksim \mathcal{N} \left( \xi_{noBAO,\theta}, C \right)$  & \textbf{- Generalized Likelihood ratio} &  \textbf{- Model-dependent} \\
				 & $\mathcal{H}_1:$  &  $\exists \, \theta \,\, \mbox{s.t.} \,\, \hat{\xi} \thicksim \mathcal{N} \left( \xi_{BAO,\theta}, C \right)$ 	  &   							&   \textbf{- Overestimation of significance } \\
  	& & 	&  	&   \textbf{- Constant covariance matrix } \\
    	& &  	&    	&   - Unstable for small binning \\
     	& & 	&  	&   - Gaussian hypothesis \\
      	&  &     & 	& \\
\it{$\Delta l$ method} 		&  $\mathcal{H}_0:$ &  $\exists \, \theta  \,\, \mbox{s.t.} \,\, \hat{\xi} \thicksim \mathcal{N} \left( \xi_{noBAO,\theta}, C_{noBAO,\theta} \right)$  	& \textbf{- Generalized Likelihood ratio} 	    	&   \textbf{- Model-dependent} \\
						&  $\mathcal{H}_1:$  &  $\exists \, \theta  \,\, \mbox{s.t.} \,\, \hat{\xi} \thicksim \mathcal{N} \left( \xi_{BAO,\theta}, C_{BAO,\theta}  \right)$ 	&\textbf{- Variations of covariance matrix} &  \textbf{-  Long computation time $\rightarrow$}\\
 	 & 				&				 & - Better results than $\Delta \chi^2$ statistic 	&  \;\;\;\,\textbf{imprecise for high significance} \\
   	 & 				 &   		  		 &   \;\;\;\,for varying covariance matrix	&   - Unstable for small binning \\
    	 & & &  &   - Gaussian hypothesis \\
	 & & &  &   \\
\tableskip\tableline\tableline\tableskip
\end{tabular}
\end{center}
NOTES.---%
The most important points are in bold. We found that the Gaussian hypothesis is well verified in practice, and that using large bins is not a serious problem. A major difference is whether methods give correct estimate of the significance. Other important differences come from the tested hypotheses $\mathcal{H}_0$ and $\mathcal{H}_1$: whether they are based on a full modeling of $\hat{\xi}$, whether they allow variations of the covariance matrix.
\end{table*}

\newpage
\appendix

\section{Best-fit $\chi^2$}
\label{best_fit}
We consider a class of binned model correlation functions, $\xi_\theta=(\xi_\theta(r_i))_{1\le i \le n}$, with a $k$-dimensional parameter $\theta=(\theta_1, \dots, \theta_k)\in \Theta$. We suppose that the estimator $\hat{\xi}$ of the correlation function is Gaussian with covariance matrix $C$ and expectation inside the model space (i.e. $\exists \, \theta_0$ such that $\hat{\xi} \thicksim \mathcal{N}\left(\xi_{\theta_0},C \right)$). We look at the $\chi_\theta^2$ statistic which has a dependence on $\theta$

\begin{equation}
\chi_\theta^2 = \sum_{1\leq i,j\leq n}  \left[ \hat{\xi}(r_i)-\xi_\theta(r_i) \right] C^{-1}_{i,j}  \left[ \hat{\xi}(r_j)-\xi_\theta( r_j) \right] 
\end{equation}

Now we make the important assumption that the space of model correlation function $(\xi_\theta)_{\theta \in \Theta}$ is a $k$-dimensional affine subspace of $\mathbb{R}^n$. Then the best-fit $\chi_\theta^2$ value follows a chi-square distribution with a number of degrees of freedom equal to $n-k$, i.e. the measurement dimension minus the parameter space dimension

\begin{equation}
\min_{\theta} \chi_\theta^2 \thicksim \chi_{n-k}^2
\end{equation}

Since $C$ is a positive definite matrix, we can consider $C^{-1/2}$. Let us note $\hat{X}=C^{-1/2} (\hat{\xi}-\xi_{\theta_0})$ and $X_\theta=C^{-1/2} (\xi_\theta-\xi_{\theta_0})$, so that we can rewrite the $\chi_\theta^2$ statistic as

\begin{equation}
\chi_\theta^2 = \| \hat{X}- X_\theta \|^2
\end{equation}

This is the Karhunen-Lo\`eve transform which consists in whitening the measurement vector $\hat{\xi}$. This means that the resulting vector $\hat{X}$ is a multivariate Gaussian variable with expected value 0 and covariance matrix equal to the identity. Indeed the covariance matrix of $\hat{X}$ is equal to

\begin{eqnarray*}
\mathbb{E}[ \hat{X} \hat{X}^T] &=& C^{-1/2}   \mathbb{E}[ (\hat{\xi}-\xi_{\theta_0}) (\hat{\xi}-\xi_{\theta_0})^T] C^{-1/2}  \\
& = & C^{-1/2}  C C^{-1/2} = I_n 
\end{eqnarray*}

with $I_n$ the $n$ x $n$ identity matrix. Thus, in any orthonormal basis of $\mathbb{R}^n$, the $n$ components of $\hat{X}$ are independent standard Gaussian variables. Let us write $F_\Theta=(X_\theta)_{\theta \in \Theta}$, which is a $k$-dimensional vectorial space, and $F^\perp_\Theta$ its orthogonal complement of dimension $n-k$. Let us write $(\hat{Y}_1, \dots, \hat{Y}_{k+1}, \dots, \hat{Y}_n)$ the components of $\hat{X}$ into an orthonormal basis, which has the first $k$ vectors in $F_\Theta$ and the last $n-k$ vectors in $F^\perp_\Theta$. Then the $\hat{Y}_i$'s are independent standard normal variables. Moreover $\chi_\theta^2$ is minimized when $X_\theta$ is the projection of $\hat{X}$ onto $F_\Theta$, and equals

\begin{eqnarray}
\min_\theta \chi^2_\theta  & = &  \| \hat{X}- \hat{X}_{F_\Theta} \|^2 
					=  \| \hat{X}_{F^\perp_\Theta} \|^2 \\
					& = &  \sum^n_{i=k+1} \hat{Y}^2_i
\end{eqnarray}

This shows that the best-fit $\chi_\theta^2$ follows a chi-square distribution with $n-k$ degrees of freedom, i.e. $\min_\theta \chi^2_\theta \thicksim \chi^2_{n-k}$.


\section{Difference of best-fits $\chi^2$ in nested models}
\label{nested_best_fit}
Here we consider two nested classes of model correlation functions, $\xi_\theta$ with $\theta \in \Theta_1$ for the first class and  $\theta \in \Theta_2$ for the second class. We suppose that $\Theta_1$ is $k$-dimensional and that $\Theta_2$ is $(k+l)$-dimensional with $\Theta_1 \subset \Theta_2$.

We still suppose that the estimator $\hat{\xi}$ is Gaussian with covariance matrix $C$ and expectation inside the restricted class (i.e. $\exists \, \theta_0 \in \Theta_1$ such that $\hat{\xi} \thicksim \mathcal{N}\left(\xi_{\theta_0},C \right)$). We also keep the assumption that the spaces of model correlation functions $(\xi_\theta)_{\theta \in \Theta_1}$ and  $(\xi_\theta)_{\theta \in \Theta_2}$ are affine subspaces of $\mathbb{R}^n$ of respective dimensions $k$ and $k+l$. Then the difference of best-fits between the two classes $\Theta_1$ and $\Theta_2$ follows a chi-square distribution with number of degrees of freedom equal to $l$, i.e. the difference in the number of parameters of the two classes.
\begin{equation}
\min_{\theta \in \Theta_1} \chi_\theta^2 - \min_{\theta \in \Theta_2} \chi_\theta^2  \thicksim \chi_{l}^2
\end{equation}

This follows easily from appendix \ref{best_fit}. We consider again the Karhunen-Lo\`eve transforms $\hat{X}=C^{-1/2} (\hat{\xi}-\xi_{\theta_0})$ and $X_\theta=C^{-1/2} (\xi_\theta-\xi_{\theta_0})$. Let us write the model spaces $F_{\Theta_1}=(X_\theta)_{\theta \in \Theta_1}$ and $F_{\Theta_2}=(X_\theta)_{\theta \in \Theta_2}$, and their orthogonal complements $F^\perp_{\Theta_1}$ and $F^\perp_{\Theta_2}$. We can write $(\hat{Y}_1, \dots, \hat{Y}_{k+1}, \dots, \hat{Y}_{k+l+1}, \dots \hat{Y}_n)$ the components of $\hat{X}$  into an orthonormal basis, which has the first $k$ vectors in $F_{\Theta_1} \cap F_{\Theta_2}$, the next $l$ components in $F^\perp_{\Theta_1} \cap  F_{\Theta_2}$ and the last $n-(k+l)$ components in $F^\perp_{\Theta_1}  \cap F^\perp_{\Theta_2}$. The components $\hat{Y}_i$'s are independent standard normal variables. Moreover for each class of model, $\chi_\theta^2$ is minimized when $X_\theta$ is the projection of $\hat{X}$ onto the model space $F_\Theta$

\begin{eqnarray*}
\min_{\theta \in \Theta_1} \chi^2_\theta & =  & \| X_{F^\perp_{\Theta_1}} \|^2  =  \sum^n_{i=k+1} \hat{Y}^2_i \\
\min_{\theta \in \Theta_2} \chi^2_\theta & =  & \| X_{F^\perp_{\Theta_2}} \|^2  =  \sum^n_{i=k+l+1} \hat{Y}^2_i 
\end{eqnarray*}

So the best-fit difference is given by
\begin{equation}
\min_{\theta \in \Theta_1} \chi_\theta^2 - \min_{\theta \in \Theta_2} \chi_\theta^2  = \sum^{k+l}_{i=k+1} \hat{Y}^2_i 
\end{equation}

This shows that the difference of best-fits $\chi_\theta^2$ follows a chi-square distribution with $l$ degrees of freedom, i.e. $\min_{\theta \in \Theta_1} \chi_\theta^2 - \min_{\theta \in \Theta_2} \chi_\theta^2 \thicksim \chi^2_l$.


\section{Optimality of the likelihood ratio}
\label{optimality_lratio}
Let us consider the likelihood ratio $\Lambda(\hat{\xi})=\mathcal{L}_{\mathcal{H}_0}(\hat{\xi}) / \mathcal{L}_{\mathcal{H}_1}(\hat{\xi})$ and another statistic $S(\hat{\xi})$ for testing the hypotheses $\mathcal{H}_0$ and $\mathcal{H}_1$. Let us suppose that $\mathcal{H}_1$ is preferred over $\mathcal{H}_0$ for low values of $S(\hat{\xi})$ as for the likelihood ratio (if this is not the case we just consider $-S$).

We first consider statistical tests for a given significance of $\alpha$. The test based on the likelihood ratio is

\begin{itemize}
\item if $\Lambda(\hat{\xi}) \leq \eta_\Lambda$ then accept $\mathcal{H}_1$ 
\item if $\Lambda(\hat{\xi}) > \eta_\Lambda$ then accept $\mathcal{H}_0$
\end{itemize}

The test based on the statistic $S$ is
\begin{itemize}
\item if $S(\hat{\xi}) \leq \eta_S$ then accept $\mathcal{H}_1$ 
\item if $S(\hat{\xi}) > \eta_S$ then accept $\mathcal{H}_0$
\end{itemize}

with $\alpha=P\left(\Lambda(\hat{\xi}) \leq \eta_\Lambda \, | \, \mathcal{H}_0 \right)$ and $\alpha=P\left(S(\hat{\xi}) \leq \eta_S \, | \, \mathcal{H}_0 \right)$. The Neyman-Pearson lemma states that the power of the likelihood ratio test if larger than the power of any other test. This means that the probability of accepting $\mathcal{H}_1$ if it is true is larger for the likelihood ratio test.

\begin{equation}
P\left(\Lambda(\hat{\xi}) \leq \eta_\Lambda \, | \, \mathcal{H}_1 \right) \geq P\left(S(\hat{\xi}) \leq \eta_S \, | \, \mathcal{H}_1 \right)
\label{p_ineq}
\end{equation}

Now we consider the significances corresponding to realization values $\Lambda(\hat{\xi})=x$ and $S(\hat{\xi})=y$ that we write respectively $\alpha_{\Lambda}$ and $\alpha_S$
\begin{eqnarray*}
\alpha_{\Lambda}(x) &=& P(\Lambda(\hat{\xi}) \leq x \, | \, \mathcal{H}_0) \\
\alpha_S(y) &=& P(S(\hat{\xi}) \leq y \, | \, \mathcal{H}_0)
\end{eqnarray*}

We have $\alpha_\Lambda(\eta_\Lambda)=\alpha$ and  $\alpha_S(\eta_S)=\alpha$. Since $\alpha_\Lambda$ and $\alpha_S$ are increasing functions, the conditions $\Lambda(\hat{\xi}) \leq \eta_\Lambda$ and $S(\hat{\xi}) \leq \eta_S$ are equivalent respectively to $\alpha_{\Lambda}\left (\Lambda(\hat{\xi})\right)\leq \alpha$ and $\alpha_S \left( S(\hat{\xi}) \right) \leq \alpha$. If we simplify the notations and write $\alpha_\Lambda(\hat{\xi})$ for $\alpha_{\Lambda}\left (\Lambda(\hat{\xi})\right)$ and $\alpha_S(\hat{\xi})$ for $\alpha_S \left( S(\hat{\xi}) \right)$, we obtain from equation (\ref{p_ineq}) that for any $\alpha$
\begin{equation}
P\left( \alpha_{\Lambda}(\hat{\xi}) \leq \alpha \, | \, \mathcal{H}_1 \right) \geq P\left( \alpha_S(\hat{\xi})    \leq \alpha  \, | \, \mathcal{H}_1 \right)
\label{p_ineq2}
\end{equation}

Let us show that this implies
\begin{equation}
\mathbb{E}\left[  \alpha_\Lambda(\hat{\xi}) \, | \, \mathcal{H}_1  \right] \leq \mathbb{E}\left[  \alpha_S(\hat{\xi}) \, | \, \mathcal{H}_1  \right]
\end{equation}

In what follows, we always keep the condition $\mathcal{H}_1$ in expectations and probabilities, so we omit it to simplify the notations. We write $F_\Lambda$ and $F_S$ the cumulative distribution functions given by $F_\Lambda(\alpha)=P(\alpha_\Lambda(\hat{\xi}) \leq \alpha)$ and $F_S(\alpha)=P(\alpha_S(\hat{\xi}) \leq \alpha)$. Equation (\ref{p_ineq2}) implies for any $\alpha$ and $p$
\begin{eqnarray}
\label{F_ineq1}
F_\Lambda(\alpha) \geq F_S(\alpha) \\
F^{-1}_\Lambda(p) \leq F^{-1}_S(p) 
\label{F_ineq2}
\end{eqnarray}

The expectation of $\alpha_\Lambda(\hat{\xi})$ is given by
\begin{equation}
\mathbb{E}[\alpha_\Lambda(\hat{\xi})]=\int \alpha \, dF_\Lambda(\alpha) = \int_0^1 F^{-1}_\Lambda(p)\, dp
\end{equation}

where we made the change of variable $p=F_\Lambda(\alpha)$. The same computation can be made for $\mathbb{E}[\alpha_S(\hat{\xi})]$, and since $F^{-1}_\Lambda(p) \leq F^{-1}_S(p) $ we get $\mathbb{E}[\alpha_\Lambda(\hat{\xi})] \leq \mathbb{E}[\alpha_S(\hat{\xi})]$. So the expected significance given as a $p$-value is minimized for the likelihood ratio. 

If we measure the significance as a number of $\sigma$ instead of a $p$-value, both inequalities (\ref{F_ineq1}) and (\ref{F_ineq2}) are reversed. So the inequality on the expected values is also reversed, and the expected number of $\sigma$ is maximized for the likelihood ratio.

\end{document}